\documentclass[twocolumn]{jpsj2}
\usepackage{graphicx}
\usepackage{amsmath,amssymb}

\def\runauthor{Y. Yanase, M. Mochizuki and M. Ogata}

\title{ 
Role of Spin-Orbit Coupling on the Spin Triplet Pairing in 
${\rm Na_{x}Co_{}O_{2}} \cdot y{\rm H}_{2}{\rm O}$ 
\\ 
II: Multiple phase diagram under the Magnetic Field
} 

\author{Youichi {\sc Yanase}\footnote{E-mail:
yanase@hosi.phys.s.u-tokyo.ac.jp}, Masahito {\sc Mochizuki} 
and Masao {\sc Ogata}}

\inst{Department of Physics, University of Tokyo, Tokyo 113-0033}

\recdate{Today 2005}

\abst
{
 The possibility of multiple phase diagram in the novel superconductor 
${\rm Na_{x}Co_{}O_{2}} \cdot y{\rm H}_{2}{\rm O}$ is analyzed 
on the basis of the multi-orbital Hubbard model including 
the atomic spin-orbit coupling. 
 We have shown that the spin triplet pairing state is stable in this model. 
 The $p$-wave ($f$-wave) state is stabilized when the Hund's rule 
coupling is large (small). 
 In the precedent paper, we have determined the direction of $d$-vector 
at $T=T_{\rm c}$ and $H=0$ within the linearized Dyson-Gorkov equation. 
 In this paper, the pairing state below $T_{\rm c}$ and under the 
magnetic field is determined within the weak coupling approximation 
including the paramagnetic effect. 
 We find that the $p+f$ coexistent state is stabilized at low 
temperatures in a part of parameter range. 
 We point out that the phase diagram in the $H$-$T$ plane 
is quite different between the $p$-wave, $f$-wave and $p+f$-wave 
superconductivities. 
 The characteristics of each phase are clarified 
by showing the magnetic susceptibility and specific heat. 
 We discuss the comparison with experimental results and suggest 
some future experiments to detect the multiple phase transition. 
}

\kword
{
${\rm Na_{x}Co_{}O_{2}} \cdot y{\rm H}_{2}{\rm O}$; 
multi-orbital model;
spin triplet superconductivity; 
$d$-vector; 
multiple phase diagram
}

\begin{document}
\sloppy
\maketitle

\newcommand{\eli}{$\acute{{\rm E}}$liashberg }
\renewcommand{\k}{\mbox{\boldmath$k$}}
\newcommand{\q}{\mbox{\boldmath$q$}}
\newcommand{\Q}{\mbox{\boldmath$Q$}}
\newcommand{\kk}{\mbox{\boldmath$k'$}}
\newcommand{\e}{\varepsilon}
\newcommand{\ee}{\varepsilon^{'}}
\newcommand{\s}{{\mit{\it \Sigma}}}
\newcommand{\J}{\mbox{\boldmath$J$}}
\newcommand{\vv}{\mbox{\boldmath$v$}}
\newcommand{\Jh}{J_{{\rm H}}}
\newcommand{\LL}{\mbox{\boldmath$L$}}
\renewcommand{\SS}{\mbox{\boldmath$S$}}
\newcommand{\Tc}{$T_{\rm c}$ }
\newcommand{\Tcf}{$T_{\rm c}$}
\newcommand{\Co}{${\rm Na_{x}Co_{}O_{2}} \cdot y{\rm H}_{2}{\rm O}$ }
\newcommand{\Cof}{${\rm Na_{x}Co_{}O_{2}} \cdot y{\rm H}_{2}{\rm O}$}
\newcommand{\tgf}{$t_{\rm 2g}$-orbitals}
\newcommand{\tg}{$t_{\rm 2g}$-orbitals }
\newcommand{\av}{\mbox{\boldmath${\rm a}$} }
\newcommand{\bv}{\mbox{\boldmath${\rm b}$} }
\newcommand{\avf}{\mbox{\boldmath${\rm a}$}}
\newcommand{\bvf}{\mbox{\boldmath${\rm b}$}}
\newcommand{\egf}{$e_{\rm g}$-Fermi surface }
\newcommand{\egff}{$e_{\rm g}$-Fermi surface}
\newcommand{\agf}{$a_{\rm 1g}$-Fermi surface }
\newcommand{\agff}{$a_{\rm 1g}$-Fermi surface}

\section{Introduction}

 The spin triplet superconductivity is one of the most exciting topics 
in condensed matter physics. 
 Vast studies have been devoted to the candidate materials such as 
Sr$_2$RuO$_4$,~\cite{rf:maeno} (TMTSF)$_2$PF$_6$,~\cite{rf:TMTSF} 
UPt$_3$,~\cite{rf:tou} 
UNi$_2$Al$_3$~\cite{rf:UNiAl}, UGe$_2$~\cite{rf:saxena} and 
URhGe~\cite{rf:aoki}. 
 Recently, a novel superconductivity has been discovered in 
\Co~\cite{rf:takada} and the possibility of spin triplet 
superconductivity has attracted huge interests.

 The properties of superconducting state in \Co have been 
investigated by many experimental studies including the 
magnetization,~\cite{rf:sakurai} 
NMR,~\cite{rf:yoshimura,rf:kobayashi,rf:zheng,rf:ishida,rf:ihara-o}
$\mu$SR,~\cite{rf:higemoto,rf:uemura,rf:kanigel}
specific heat,~\cite{rf:hdyang,rf:lorenz,rf:oeschler}
resistivity,~\cite{rf:sasaki} and impurity effect.~\cite{rf:yokoi} 
 Except for the impurity effect,~\cite{rf:yokoi} 
most of them indicate a non-$s$-wave pairing. 
 The discovery of magnetic phase in the neighborhood of 
superconducting phase~\cite{rf:ihara,rf:sakuraimagnetic} 
indicates a strong electron correlation which generally 
leads to an unconventional superconductivity.

 As summarized in Ref.~24, there have been many theoretical 
studies.~\cite{rf:baskaran,rf:shastry,rf:lee,rf:ogata,rf:ikeda,
rf:kuroki,rf:Ytanaka,rf:motrunich,rf:kurokii-wave,rf:yata,rf:khaliullin}
 Although most of these studies have assumed single-orbital models, 
the justification of this assumption is not clear because 
the conduction band in \Co has orbital degeneracy, as pointed out by 
Koshibae {\it et al.}~\cite{rf:koshibae} 
 In order to examine the superconductivity in the multi-orbital system, 
we have constructed a three-orbital Hubbard model~\cite{rf:mochizuki} 
which appropriately reproduces the electronic structure obtained 
in the LDA calculations.~\cite{rf:singh,rf:pickett,rf:arita} 
 From the results of perturbation theory~\cite{rf:yanase} 
and FLEX approximation,~\cite{rf:mochizuki} 
we have shown that the spin triplet superconductivity is stable 
in the wide parameter range. 
 Then, the $p$-wave superconductivity and $f$-wave superconductivity 
are nearly degenerate because of the character of orbital in each 
Fermi surface.~\cite{rf:yanase} 
 We have found an orbital dependent superconductivity like 
Sr$_2$RuO$_4$~\cite{rf:agterberg} and derived the two-orbital 
Hubbard model reproducing the $e'_{\rm g}$-doublet which mainly 
leads to the superconductivity.

 In the precedent paper,~\cite{rf:yanasepart1} which we call ``I'' 
in the following, we have constructed a 
two-orbital Hubbard model including the spin-orbit coupling term 
and determined the $d$-vector at $T=T_{\rm c}$ with use of the 
linearized Dyson-Gorkov equation.
 In the present paper, we study the phase diagram in the temperature 
and the magnetic field plane ($H$-$T$ plane) and clarify the physical 
properties, such as magnetic susceptibility and specific heat, in 
each phase. This study is for the following two purposes. 
 (1) One purpose is to provide clear physical quantities which can be 
compared in experiments. 
 Many measurements including the NMR Knight shift are performed 
under the magnetic field. 
 Since the $d$-vector may rotate under the magnetic field 
so as to gain the Zeeman coupling energy, it is highly desired 
to understand the $d$-vector under the magnetic field. 
 (2) Another and more interesting purpose is to suggest a 
possibility of multiple phase transition in the superconducting state. 
 The multiple phase diagram may appear in the $H$-$T$ plane 
owing to the rotation of $d$-vector.  
 If the multiple phase transition is observed in experiments, 
that will be a strong evidence for the spin triplet pairing.

 The effects of spin-orbit coupling ware clarified in I. 
 In the model without any spin-orbit coupling term, 
the $d$-vector rotates in the infinitesimal magnetic field.  
 In general, the $d$-vector is determined by the competition between 
the magnetic field and the anisotropy arising from the spin-orbit 
coupling. 
 For superconductors, it is reasonable to consider the atomic 
spin-orbit coupling, namely so-called $L$-$S$ coupling as a 
microscopic origin of the spin-orbit coupling for Cooper pairs. 
 Therefore, the $d$-vector in spin triplet superconductors is a 
fundamental issue of multi-orbital systems.

 The multiple phase transition in heavy fermion superconductors has 
attracted much interests for last two 
decades.~\cite{rf:machida,rf:sauls,rf:joynt,rf:sigrist} 
 Then, the theoretical treatment has relied on the phenomenological 
theory and the role of $L$-$S$ coupling on the $d$-vector has been 
a longstanding problem.  
 For instance, the pairing symmetry in UPt$_3$ has been discussed for a 
long time and the anisotropy of $d$-vector is still under 
debate.~\cite{rf:machida,rf:sauls,rf:joynt} 
 We consider that the study on the $d$-electron systems will be a first 
step for this problem. 
 The microscopic theory including the spin-orbit coupling can be applied 
to the $d$-electron systems like \Co and Sr$_2$RuO$_4$ because 
these systems have a simple electronic structure in comparison with 
$f$-electron systems. 
 Actually, we have performed a microscopic analysis on the $d$-vector 
in Sr$_2$RuO$_4$~\cite{rf:yanaseRuSO,rf:yanasereview} and obtained 
the consistent results with experiments.~\cite{rf:murakawa,rf:luke} 
 We have found that Sr$_2$RuO$_4$ is a particular case in the sense 
that the anisotropy of $d$-vector is very small owing to 
the orbital dependent superconductivity. This result has been 
confirmed by the NMR measurement.~\cite{rf:murakawa} 
 In this paper, we show that \Co is a more general case where 
the spin-orbit coupling plays a more important role. 
 We point out that the multiple phase transition will occur 
when the order parameter includes the $p$-wave component.

 This paper is organized as follows. 
 In \S2, we briefly summarize the results of linearized Dyson-Gorkov 
equation for the two-orbital Hubbard model and derive an effective 
single-band model including the Zeeman coupling term. 
 In \S3, we determine the phase diagram in $H$-$T$ plane 
within the mean field theory. 
 We obtain three pairing states at zero magnetic field, 
namely the $p$-wave, $f$-wave and $p+f$-wave states. 
 The $d$-vector under the magnetic field are investigated 
for these states in \S3.1, \S3.2 and \S3.3, respectively. 
 Our interests are focused on the magnetic field parallel to the 
two-dimensional plane because a wide variety of phase diagram is
expected in this direction and also because the Knight shift has
been measured under the parallel magnetic field 
in most cases.~\cite{rf:yoshimura,rf:kobayashi,rf:ihara-o,rf:ishidaprivate,
rf:zhengprivate,rf:higemoto} 
 The nature of multiple phase transition is analyzed with attention 
to the thermal and magnetic properties. 
 In \S4, the role of vortex state and that of $a_{\rm 1g}$-orbital 
which are ignored in \S3 will be discussed. 
 We show that an extra phase transition may be induced by these 
effects. 
 Some discussions are given in the last section \S5.

\section{Formulation}

\subsection{Pairing state at $T=T_{\rm c}$ and $H=0$}

 As in I, we adopt a two-orbital Hubbard model reproducing 
the $e'_{\rm g}$-doublet. 
 From the analysis of three orbital Hubbard model without including 
the spin-orbit coupling, we have found that the superconductivity 
is mainly induced by the $e'_{\rm g}$-orbitals.~\cite{rf:mochizuki,rf:yanase}
 Indeed, the two-orbital Hubbard model without including the 
$a_{\rm 1g}$-orbital appropriately reproduces the results of 
three-orbital Hubbard model.~\cite{rf:yanase} 
 We choose the parameters of two-orbital Hubbard model as in 
I ~\cite{rf:yanasepart1} unless we specify.

 The pairing symmetry in the two-orbital and three-orbital Hubbard 
models are basically determined by the two parameters, namely 
$n_{\rm e}$ and $\Jh$. 
 Here, $n_{\rm e}$ is the holes in the \egf and $\Jh$ is 
the Hund's rule coupling. 
 We have found that the $f$-wave superconductivity is favored 
when $n_{\rm e}$ is large and 
{\it vice versa}.~\cite{rf:yanase,rf:yanasepart1} 
 This is because the amplitude of order parameter 
in the $p$-wave symmetry have to be small around the K-point owing to the 
periodicity of Brillouin zone. 
 As increasing $n_{\rm e}$, the \egf approaches to the K-point and the 
$p$-wave superconductivity becomes unfavorable. 
 As for $\Jh$, the $p$-wave ($f$-wave) superconductivity is stabilized 
when $\Jh$ is large (small).

 The $d$-vector is determined by the spin-orbit coupling term which 
plays an essential role for the issues in this paper. 
 The coupling constant $2 \lambda$ of spin-orbit coupling term has 
been estimated as $57$meV from NMR measurements~\cite{rf:michioka}. 
 This value corresponds to $\lambda=0.17$ in our unit if we choose 
the band width $W \sim 9 = 1.5$eV according to the LDA calculations. 
 Although this estimation has some ambiguities, the spin-orbit coupling 
in this order is much smaller than the band width. 
 This is in contrast with the RVB theory discussed 
by Khaliullin {\it et al.}~\cite{rf:khaliullin} who have considered 
the opposite limit $\lambda = \infty$.

 If the spin-orbit coupling is neglected, there remains a 
$3 \times 2=6$-fold degeneracy in the $p$-wave state, 
while the degeneracy in the $f$-wave state is $3$-fold. 
 Due to the spin-orbit coupling, the degeneracy is lifted and the 
pairing states are classified into $P_{\rm xy+}$, $P_{\rm xy-}$ and 
$P_{\rm z}$ for the $p$-wave symmetry while 
$F_{\rm xy}$ and $F_{\rm z}$ for the $f$-wave symmetry. 
 The $d$-vector in these states has been summarized in Table II 
of I.~\cite{rf:yanasepart1} 
 We have determined the pairing state at $T=T_{\rm c}$ and $H=0$ 
with use of the linearized Dyson-Gorkov equation 
whose derivation for the $SU(2)$ asymmetric system has been given 
in I and Ref.~47. 
 For the two-orbital Hubbard model, we have obtained the two independent 
equations as, 
\begin{eqnarray}
\label{eq:eliashberg-equation-uu} 
&& \hspace{-10mm}
\lambda^{\uparrow \uparrow}_{\rm e} \Delta_{\uparrow \uparrow}(k) = 
- \sum_{k'} V_{\uparrow \uparrow}(k,k') |G_{2}(k')|^{2} 
\Delta_{\uparrow \uparrow}(k'), 
\\
\label{eq:eliashberg-equation-ud} 
&& \hspace{-10mm}
\lambda^{\uparrow \downarrow}_{\rm e} \Delta_{\uparrow \downarrow}(k) = 
- \sum_{k'} V_{\uparrow \downarrow}(k,k') |G_{2}(k')|^{2} 
\Delta_{\uparrow \downarrow}(k'). 
\end{eqnarray}
 Here, $\Delta_{\uparrow \uparrow}$ and $\Delta_{\uparrow \downarrow}$ 
are the order parameters for the equal spin pairing state and opposite spin 
pairing state, respectively. 
 For the effective interactions, $V_{\uparrow \uparrow}(k,k')$ and 
$V_{\uparrow \downarrow}(k,k')$, we applied the second order 
perturbation theory (SOP) with respect to the Coulomb interactions, 
{\it i.e.} $U$, $U'$, $\Jh$ and $J$. 
 Note that we have confirmed that the SOP is consistent with the renormalized 
third order perturbation theory as well as the FLEX approximation 
which include higher order corrections.~\cite{rf:yanase,rf:mochizuki} 
 The Green function $G_{2}(k)$ is defined as 
$G_{2}(k)=\frac{1}{{\rm i}\omega_{n} - 
E_{2}(\mbox{{\scriptsize \boldmath$k$}})}$ where $E_{2}(\k)$ is the 
upper band composing the Fermi surface. 
 Note that $E_{2}(\k)$ is slightly affected by the spin-orbit coupling.

 By solving eqs.~(\ref{eq:eliashberg-equation-uu}) and 
(\ref{eq:eliashberg-equation-ud}), we have found that 
the spin-orbit coupling stabilizes the $P_{\rm xy+}$-state 
when the pairing symmetry is $p$-wave. 
 Then, the pairing state is described by the $d$-vector as, 
$\hat{d}=p_{\rm x}\hat{x}+p_{\rm y}\hat{y}$ or 
$\hat{d}=p_{\rm y}\hat{x}-p_{\rm x}\hat{y}$.  
 There is no violation of time-reversal-symmetry in this state
which is consistent with the $\mu$SR measurement.~\cite{rf:higemoto} 
 In the case of $f$-wave symmetry, $F_{\rm xy}$-state 
($\hat{d}=f_1 \hat{x} - \alpha f_{2} \hat{y}$ or 
$\hat{d}=\alpha f_2 \hat{x} + f_{1} \hat{y}$)
and $F_{\rm z}$-state ($\hat{d}=f_1 \hat{z}$) can be stabilized 
depending on $n_{\rm e}$ and $\Jh$.

 As shown in I, the anisotropy of $d$-vector defined as 
$|\lambda^{\uparrow \uparrow}_{\rm e}-
\lambda^{\uparrow \downarrow}_{\rm e}|/
\lambda^{\uparrow \uparrow}_{\rm e}$ is much larger in the $p$-wave 
state than in the $f$-wave state. 
 This is because the first order term in spin-orbit coupling vanishes 
in the latter case. 
 This means that the $d$-vector in the $f$-wave superconductivity 
is rotated by a small magnetic field parallel to the $d$-vector. 
 Thus, the role of spin-orbit coupling is quite different between the 
$p$-wave and $f$-wave superconductivities. 
 This leads to the qualitatively different phase diagram under 
the magnetic field as shown in \S3.1-3.

\subsection{Weak coupling approximation}

 In order to determine the pairing state below \Tc and under the 
magnetic field, we introduce an effective single-band model 
which reproduces the results of linearized Dyson-Gorkov equation 
at $H=0$. 
\begin{eqnarray}
&& \hspace{-10mm}
H_{1} = \sum_{k,s} E_{2}(\k)
c_{k,s}^{\dag} c_{k,s}
\nonumber 
\\
&& \hspace*{0mm} 
- \frac{1}{2} g_{1} \sum_{k,k',s} \phi_{1{\rm x}}(\k) \phi_{1{\rm x}}(\k') 
c_{k,s}^{\dag} c_{-k,s}^{\dag} c_{-k',s} c_{k',s}
\nonumber 
\\
&& \hspace*{0mm} 
- \frac{1}{2} g_{2} \sum_{k,k',s} \phi_{1{\rm y}}(\k) \phi_{1{\rm y}}(\k') 
c_{k,s}^{\dag} c_{-k,s}^{\dag} c_{-k',s} c_{k',s}
\nonumber 
\\
&& \hspace*{0mm} 
- \frac{1}{2} g_{3} \sum_{k,k',s}
[{\rm i} s 
\phi_{1{\rm x}}(\k) \phi_{1{\rm y}}(\k') 
c_{k,s}^{\dag} c_{-k,s}^{\dag} c_{-k',s} c_{k',s}
\nonumber
\\
&& \hspace*{15mm} -{\rm i} s 
\phi_{1{\rm y}}(\k) \phi_{1{\rm x}}(\k') 
c_{k,s}^{\dag} c_{-k,s}^{\dag} c_{-k',s} c_{k',s}
]
\nonumber 
\\
&& \hspace*{0mm} - \frac{1}{2} g_{4} 
\sum_{k,k',s} \phi_{2{\rm x}}(\k) \phi_{2{\rm x}}(\k') 
c_{k,s}^{\dag} c_{-k,-s}^{\dag} c_{-k',-s} c_{k',s}
\nonumber 
\\
&& \hspace*{0mm} - \frac{1}{2} g_{5} 
\sum_{k,k',s} \phi_{2{\rm y}}(\k) \phi_{2{\rm y}}(\k') 
c_{k,s}^{\dag} c_{-k,-s}^{\dag} c_{-k',-s} c_{k',s}
\nonumber 
\\
&& \hspace*{0mm}
- \mbox{\boldmath$M$} \mbox{\boldmath$H$}, 
\label{eq:single-orbital-model}
\end{eqnarray}
where $c_{k,s}^{\dag}$ is the creation operator for the $E_{2}(\k)$-band  
electron with pseudospin $s$. 
 The coupling constants $g_i$ are determined so as to reproduce the 
result of linearized Dyson-Gorkov equation, 
namely eqs.~(\ref{eq:eliashberg-equation-uu}) and 
(\ref{eq:eliashberg-equation-ud}). Therefore, these parameters are 
determined by the microscopic parameters of Hubbard model such as 
$n_{\rm e}$ and $\Jh$. 
 Here, $\phi_{1{\rm x}}(\k)$ and $\phi_{1{\rm y}}(\k)$ 
are the orbital part of Cooper pair wave function 
for the $d$-vector along the plane which are determined by the 
linearized Dyson-Gorkov equation in 
eq.~(\ref{eq:eliashberg-equation-uu}). 
 For instance, we obtain  
$\Delta_{\uparrow \uparrow}(\k,{\rm i}\pi T)= 
-\phi_{1{\rm x}}(\k) + {\rm i} \phi_{1{\rm y}}(\k)$ when the 
$P_{\rm xy+}$-state is stabilized at $T=T_{\rm c}$. 
 Similarly, $\phi_{2{\rm x}}(\k)$ and $\phi_{2{\rm y}}(\k)$ are 
the wave function for the $d$-vector along {\it z}-axis determined by 
eq.~(\ref{eq:eliashberg-equation-ud}). 
 In case of the $p$-wave superconductivity, $\phi_{i{\rm x}}(\k)$ and 
$\phi_{i{\rm y}}(\k)$ ($i = 1, 2$) denote the wave functions in 
the $p_{\rm x}$- and $p_{\rm y}$-wave symmetry, respectively. 
 In case of the $f$-wave superconductivity, these are wave functions 
in the $f_2$- and $f_1$-wave symmetry, respectively. 
 We have neglected the frequency dependence of $\phi_{i\alpha}(\k)$ 
for simplicity and use the value at $\omega_{n}={\rm i}\pi T$, 
{\it i.e.} the smallest Matsubara frequency.

 As a role of the magnetic field, we take into account 
the Zeeman coupling term, namely the last term 
in eq.~(\ref{eq:single-orbital-model}). 
 This term plays an essential role for the rotation of $d$-vector. 
 We obtain the magnetic moment from the two-orbital Hubbard model 
as $M_{\alpha}=(L_{\alpha}+2S_{\alpha})\mu_{\rm B}=
\sum_{\k} g_{\alpha}(\k) \tilde{S}_{\alpha}(\k) \mu_{\rm B}$ where
$\tilde{S}_{\alpha}$ is the operator of pseudospin for the 
$E_{2}(\k)$-band electron. 
 Here, the momentum dependent $g$-factor is obtained as 
$g_{\alpha}(\k)=2<\k,\tilde{S}_{\alpha}=
\frac{1}{2}|L_{\alpha}+2S_{\alpha}|\k,\tilde{S}_{\alpha}=\frac{1}{2}>$ 
where $|\k,\tilde{S}_{\alpha}=\frac{1}{2}>$ is the wave function of 
$E_{2}(\k)$-band with $\tilde{S}_{\alpha}=\frac{1}{2}$. 
 Note that $\alpha={\rm x},{\rm y}$ and ${\rm z}$ is the symmetry axis 
of crystal and $g_{\rm x}(\k)=g_{\rm y}(\k) \ne g_{\rm z}(\k)$. 
 In the following, we consider the magnetic field along {\it x}-axis 
unless we mention.

 In the following calculation, we simply ignore the orbital effect 
which induces the vortex state. 
 We discuss the orbital effect in \S4.1 and show that the 
phase diagram in the high field region $H \sim H_{\rm c2}$ can be affected 
when the orbital part of order parameter has multi-component as 
in the $p$-wave superconductivity.

 Because we have assumed separable pairing interactions, 
the following two effects are ignored in 
eq.~(\ref{eq:single-orbital-model}). 
 First, although the momentum dependence of wave functions 
$\phi_{i\alpha}(\k)$ ($i=1,2$ and $\alpha={\rm x},{\rm y}$) depends 
on the temperature and magnetic field in general,~\cite{rf:sigrist} 
we have ignored this deformation of $\phi_{i\alpha}(\k)$. 
 Second, the feedback effect, namely the effect of superconductivity 
on the effective interaction $V_{s s'}(k,k')$ is neglected. 
 Although a part of the feedback effect can be represented by 
the temperature dependence of attractive interaction, we have 
introduced temperature independent coupling constants $g_{i}$ in  
eq.~(\ref{eq:single-orbital-model}). 
 These simplifications are justified in the weak coupling region 
$T_{\rm c}/W \ll 1$ which is reasonable in \Cof. 
 Qualitatively the same assumptions have been adopted 
in the analysis of Sr$_2$RuO$_4$.~\cite{rf:nomuragap}

 In principle, these effects can be taken into account in the non-linear 
Dyson-Gorkov equation for the two-orbital Hubbard model including the 
Zeeman coupling term. 
 However, the small value of transition temperature $T_{\rm c} \sim 5$K 
makes the numerical calculation very difficult. 
 Note that we have to calculate at $T \sim 0.5$K in order to 
determine the pairing state at $T=0.1T_{\rm c}$. 
 The extraordinary large size simulation is needed to 
discuss such a small energy scale. 
 If a larger value of $T_{\rm c}$ than $5$K is assumed 
to enable the numerical calculation, we will overestimate 
contributions ignored in the weak coupling theory. 
 For instance, the renormalization of effective interaction due to 
the superconducting order parameter is overestimated. 
 As for the effect of magnetic field, the contributions scaled by  
$\mu_{\rm B}H/W$ are also overestimated, although only the contributions 
scaled by $\mu_{\rm B}H/T_{\rm c}$ are taken into account in the 
weak coupling limit $T_{\rm c}/W \rightarrow 0$. 
 It is reasonable to consider that \Co is in the weak coupling region  
$T_{\rm c}/W \ll 0$ since $T_{\rm c} \sim 5$K. 
 Therefore, we take the weak coupling limit from the beginning and 
solve the weak coupling model in eq.~(\ref{eq:single-orbital-model}) 
within the mean field theory.

 In the following, we solve the mean field equation for $H_{1}$ 
and find out the 
local minimum and saddle point of the free energy by performing the 
stability analysis. 
 As a result, we search the minimum of the free energy 
and determine the phase transition. 
 It should be noticed that the symmetry of Hamiltonian is reduced by 
the parallel magnetic field which we focus on. 
 Then, we can only rely on the $U(1)$ gauge symmetry and there are 
$6 \times 2 -1=11$ independent order parameters including the relative 
phase.

\begin{figure}[htbp]
\begin{center}
\includegraphics[width=7.5cm]{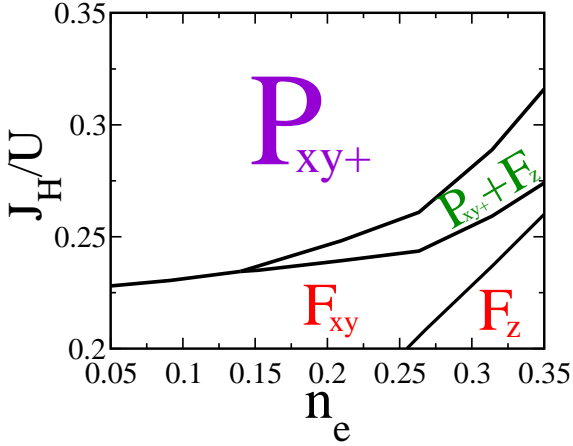}
\caption{
 Pairing state at $H=0$ and $T = 0.1 T_{\rm c}^{0}$. 
 Here, $T_{\rm c}^{0}$ is the transition temperature at which 
the superconducting instability occurs, and we choose 
the parameter $U$ so that $T_{\rm c}^{0}=6$K. 
 The notations of $P_{\rm xy+}$, $F_{\rm xy}$ and $F_{\rm z}$ 
have been given in Table II in I.~\cite{rf:yanasepart1}
 In the $P_{\rm xy+}+F_{\rm z}$-state, 
order parameters in the $P_{\rm xy+}$- and $F_{\rm z}$-states 
coexist. 
} 
\label{fig:a06}
\end{center}
\end{figure}

 Before showing the result under the magnetic field, we show the 
phase diagram at $H=0$ and below \Tc in Fig.~1. 
 We see the four states including the $P_{\rm xy+}+F_{\rm z}$-state. 
 In this state, the $p$-wave order parameter with 
$d$-vector parallel to the plane coexists with the $f$-wave order 
parameter which has the $d$-vector along {\it z}-axis. 
 This coexistent state is stabilized around the phase boundary between 
the $p$-wave and $f$-wave states so as to gain the condensation 
energy by making the superconducting gap more isotropic. 
 In this paper, we denote this pairing state as $p+f$-wave state, 
for simplicity.  
 In a small part of the parameter range, another pairing state such as 
$P_{\rm z}+F_{\rm xy}$-state may be stable. 
 However, we ignore this possibility since the parameter region will be 
very small.

 As will be shown in the following sections, the pairing state 
under the magnetic field is determined by three effects. 
 (I) The spin-orbit coupling term which favors the pairing state shown 
in Fig.~1. 
 (II) The magnetic field which is unfavorable for the $d$-vector 
parallel to the field. 
 (III) The momentum dependence of the superconducting gap, {\it i.e.} 
the more isotropic superconducting gap is, the more condensation energy 
is gained below \Tcf.  
 The multiple phase diagram appears in the $H$-$T$ plane 
as a result of these competing effects.

\section{Phase Diagram under the Magnetic Field}

 In the following, we discuss the $H$-$T$ phase diagram in each case, 
{\it i.e.} the $p$-wave, $f$-wave and $p+f$-wave states. 
 Hereafter, we choose the parameter $U$ so that $T_{\rm c}^{0}=6$K 
consistent with experimental value, where $T_{\rm c}^{0}$ is the 
transition temperature of superconductivity at $H=0$. 
 Note that $T_{\rm c}^{0}$ is the maximum value of transition 
temperatures for $P_{\rm xy+}$-, $F_{\rm xy}$- and $F_{\rm z}$-states 
because the other state is not stabilized at $T=T_{\rm c}^{0}$ 
and $H=0$ in our calculation.~\cite{rf:yanasepart1}

\subsection{$P$-wave state}

 First, we discuss the $p$-wave superconductivity which is realized 
in the large $\Jh$ region. We fix the parameters of two-orbital 
Hubbard model as $\Jh/U=7/24$ and $n_{\rm e}=0.21$ where we obtain 
$T_{\rm c}^{0}=6$K by choosing $U=6.47$. 
 In the linearized Dyson-Gorkov equation for the equal spin pairing, 
(eq.~(\ref{eq:eliashberg-equation-uu})), we obtain the wave functions 
$\phi_{1{\rm x}}(\k)$ and $\phi_{1{\rm y}}(\k)$ as 
$\Delta_{\uparrow \uparrow}(\k,{\rm i} \pi T)= 
-\phi_{1{\rm x}}(\k) + {\rm i} \phi_{1{\rm y}}(\k)$ 
where $\Delta_{\uparrow \uparrow}(k)$ is the eigenfunction 
corresponding to the maximum eigenvalue 
$\lambda^{\uparrow \uparrow}_{\rm e}$. 
 In eq.~(\ref{eq:eliashberg-equation-ud}), we obtain 
$\phi_{2{\rm x}}(\k)$ and $\phi_{2{\rm y}}(\k)$ from 
doubly degenerate eigenfunctions as 
$\Delta_{\uparrow \downarrow}(\k,{\rm i} \pi T)=\phi_{2{\rm x}}(\k)$ and 
$\Delta_{\uparrow \downarrow}(\k,{\rm i} \pi T)=\phi_{2{\rm y}}(\k)$. 
 The difference between $\phi_{1{\rm x}}(\k)$ and $\phi_{2{\rm x}}(\k)$
and that between $\phi_{1{\rm y}}(\k)$ and $\phi_{2{\rm y}}(\k)$ are  
slight because $\lambda \ll W$. 
 Because the amplitude of wave functons $\phi_{i\alpha}(\k)$ is not 
determined by the linearlized Dyson-Gorkov equation, 
we impose the normalization condition for $\phi_{i\alpha}(\k)$ as, 
\begin{eqnarray}
\label{eq:normalization} 
&& \hspace{-10mm}
K=\sum_{\k} \frac{|\phi_{i\alpha}(\k)|^{2}}{2 E_{2}(\k)}
\tanh\frac{E_{2}(\k)}{2 T_{\rm c}^{0}} 
\nonumber \\
&& \hspace{25mm}
 (i=1,2 \hspace{2mm} {\rm and} \hspace{2mm} \alpha={\rm x},{\rm y}),
\end{eqnarray}
where $K$ is an arbitrary value. 
 Although the coupling constant $g_{i}$ depends on the value of $K$, 
the physical quantities are obviously invariant for the choice of $K$.

 According to the symmetry of triangular lattice, we obtain 
$g_{1}=g_{2}$ and $g_{4}=g_{5}$. 
 Then, we have three independent parameters, 
namely $g_1$, $g_3$ and $g_5$. 
 At $\lambda=0$, we obtain $g_{1}=g_{5}$ and $g_{3}=0$. 
 Thus, the role of spin-orbit coupling is represented by the two 
parameters, $g_{1}-g_{5}$ and $g_{3}$.

 The parameters $g_{1}+g_{3}$ and $g_{5}$ are determined so as to 
reproduce the transition temperatures obtained in the linearized 
Dyson-Gorkov equation. 
 From eq.~(\ref{eq:single-orbital-model}), we obtain the gap equations 
at $H=0$ for the $P_{\rm xy+}$- and $P_{\rm z}$-states as, 
\begin{eqnarray}
\label{eq:mf-pxy+} 
&& \hspace{-10mm}
1=(g_{1}+g_{3}) \sum_{\k} \frac{|\phi_{1{\rm x}}(\k)|^{2}}{2 E_{2}(\k)}
\tanh\frac{E_{2}(\k)}{2 T_{\rm c}^{({\rm xy+})}}, 
\\
\label{eq:mf-pz} 
&& \hspace{-10mm}
1=g_{5} \sum_{\k} \frac{|\phi_{2{\rm y}}(\k)|^{2}}{2 E_{2}(\k)}
\tanh\frac{E_{2}(\k)}{2 T_{\rm c}^{({\rm z})}}. 
\end{eqnarray}
 Here, $T_{\rm c}^{({\rm xy+})}$ is the transition temperature 
for the $P_{\rm xy+}$-state, which is determined by 
the criterion $\lambda^{\uparrow \uparrow}_{\rm e}=1$ at 
$T=T_{\rm c}^{({\rm xy+})}$, while $T_{\rm c}^{({\rm z})}$ 
is the transition temperature for the $P_{\rm z}$-state and 
determined by the criterion 
$\lambda^{\uparrow \downarrow}_{\rm e}=1$ at $T=T_{\rm c}^{({\rm z})}$. 
 Since $T_{\rm c}^{({\rm xy+})} > T_{\rm c}^{({\rm z})}$ at $H=0$ 
as shown in Fig.~1, we obtain $T_{\rm c}^{({\rm xy+})}=T_{\rm c}^{0}=6$K. 
 We have obtained $T_{\rm c}^{({\rm z})}=0.865 T_{\rm c}^{0}$ for 
$\lambda=0.17$ by solving eq.~(\ref{eq:eliashberg-equation-ud}). 
 Thus, $g_{1}+g_{3}$ and $g_{5}$ are determined by the 
microscopic parameters, such as $U$, $\Jh$, $n_{\rm e}$ and $\lambda$, 
through $T_{\rm c}^{({\rm xy+})}$ and $T_{\rm c}^{({\rm z})}$, 
respectively. 

 In principle, the parameter $g_{1}-g_{3}$ can be determined by the 
transition temperature for the $P_{\rm xy-}$-state, 
$T_{\rm c}^{(xy-)}$ as, 
\begin{eqnarray}
\label{eq:mf-pxy-} 
1=(g_{1}-g_{3}) \sum_{\k} \frac{|\phi_{1{\rm x}}(\k)|^{2}}{2 E_{2}(\k)}
\tanh\frac{E_{2}(\k)}{2 T_{\rm c}^{(xy-)}}. 
\end{eqnarray}
 However, it is difficult to obtain the precise value of 
$T_{\rm c}^{(xy-)}$ from eq.~(\ref{eq:eliashberg-equation-uu}) 
since $T_{\rm c}^{(xy-)} < T_{\rm c}^{(xy+)}$. 
 Therefore, there remains an arbitrary parameter 
among $g_{1}$ and $g_{3}$. 
 This difficulty can be resolved by the microscopic analysis on the 
effective interactions $V_{\uparrow \uparrow}(k,k')$ and 
$V_{\uparrow \downarrow}(k,k')$. 
 As a result of the perturbation expansion in $\lambda$, 
we have found that $|g_{1}-g_{5}|$ is much smaller than 
$|g_{3}|$.~\cite{rf:yanasepart1} 
 The former is the second order term in $\lambda$, namely 
$g_{1}-g_{5}=O(\lambda^{2})$, while the latter is the first order term, 
namely $g_{3}=O(\lambda)$. 
 Therefore, we assume $g_{1}-g_{5}=0$ for a while and we will 
show the role of finite value of $g_{1}-g_{5}$ later.

\begin{figure}[ht]
\begin{center}
\includegraphics[width=6cm]{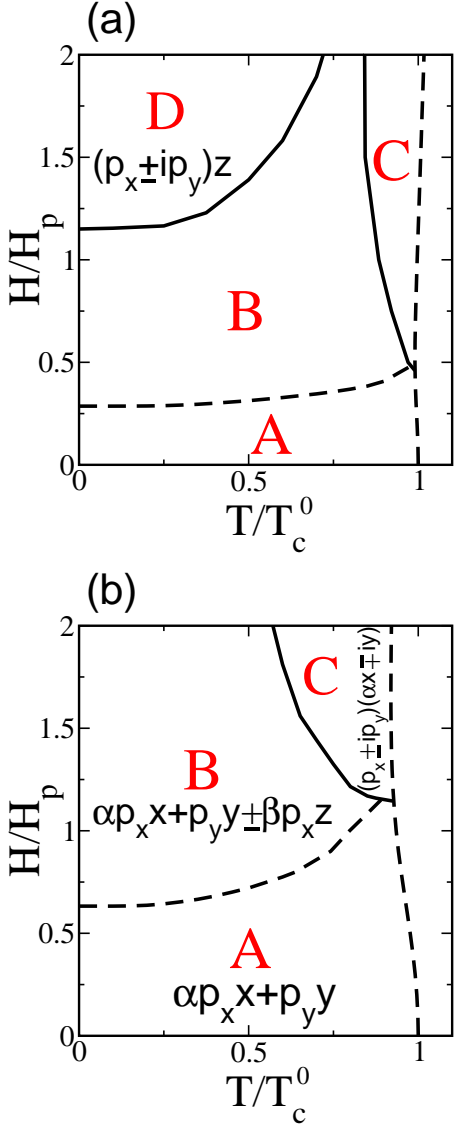}
\caption{
The phase diagram in the $p$-wave superconducting state 
under the magnetic field along {\it x}-axis. 
We choose $\lambda=0.05$ in (a) and $\lambda=0.17$ in (b). 
The notations of A, B, C and D phases are given in Table~I. 
The solid (dashed) line shows the first (second) order phase 
transition. 
} 
\end{center}
\end{figure}

 Figure 2 shows the phase diagram in $H$-$T$ plane for two values of 
$\lambda$. 
 It should be noted that the effect of magnetic field is scaled by 
the typical value $H_{\rm p}$ which is defined as 
$\mu_{\rm B}H_{\rm p}=k_{\rm B} T_{\rm c}^{0}$.  
 When we assume $T_{\rm c}^{0}=6$K, we obtain $H_{\rm p} \sim 9$T. 
 Note that $H_{\rm p}$ is a slightly smaller value than the 
Pauli paramagnetic limit. 
 We see in Fig.~2 the four phases A, B, C and D whose notations are 
given in Table~I. The coefficients $\alpha$ and $\beta$ vary according to 
the magnetic field and temperature. 
 We have not shown the small components in the order of 
$O(T_{\rm c}^{0}/W)$ which are induced by the magnetic field and 
particle-hole asymmetry.  
 Strictly speaking, A-, B-, and D-phase are the non-unitary state 
owing to these small components. 
 Recently, it has been pointed out that the non-unitarity induced by 
the paramagnetic effect plays an important role for 
Sr$_2$RuO$_4$.~\cite{rf:udagawa} 
 However, this effect is negligible in what follows since the spin-orbit
coupling is much more significant for \Co than 
Sr$_2$RuO$_4$.~\cite{rf:yanaseRuSO,rf:yanasepart1}

\begin{table}[htbp]
 \begin{center}
   \begin{tabular}{|c|c|} \hline 
A & 
$\alpha p_{\rm x}\hat{x}+p_{\rm y}\hat{y}$ 
\\\hline
B& 
$\alpha p_{\rm x}\hat{x}+p_{\rm y}\hat{y} \pm \beta p_{\rm x}\hat{z}$ 
\\\hline
C & 
$(p_{\rm x} \pm {\rm i} p_{\rm y}) (\alpha \hat{x} \mp {\rm i} \hat{y})$
\\\hline
D & 
$(p_{\rm x} \pm {\rm i} p_{\rm y}) \hat{z}$
\\\hline
E & 
$\alpha p_{\rm x}\hat{x}+p_{\rm y}\hat{y} \pm \beta f_{1} \hat{z}$ 
\\\hline
F &
$f_{1} \hat{z}$ 
\\\hline
    \end{tabular}
    \caption{
The pairing states under the magnetic field. 
The first column shows the notation in this paper. 
The second column shows the $d$-vector. 
 Note that $0 \leq \alpha, \beta \leq 1$, 
$\frac{\partial \alpha}{\partial H} < 0$ and 
$\frac{\partial \beta}{\partial H} > 0$.  
}  
  \end{center}
\end{table}

 The A-phase is connected to the $P_{\rm xy+}$-state at $H=0$ 
in an adiabatic way. 
 This phase is favored by the spin-orbit coupling and 
stabilized at low magnetic fields. 
 The $\hat{x}$-component of $d$-vector is reduced 
in the magnetic field so as to gain the Zeeman coupling energy, 
and therefore $\frac{\partial \alpha}{\partial H} < 0$. 
 Note that the A-phase has a $2$-fold degeneracy at $H=0$ or 
at $T=T_{\rm c}(H)$ where $\hat{d}=\alpha p_{\rm x}\hat{x}+p_{\rm y}\hat{y}$ 
is degenerate with $\hat{d}=\alpha p_{\rm y}\hat{x}-p_{\rm x}\hat{y}$. 
 This degeneracy is lifted for $H \ne 0$ and $T < T_{\rm c}$, and the 
former is stabilized because the order parameter in the 
$p_{\rm y}$-wave symmetry $\phi_{1{\rm y}}(\k)$ 
is more isotropic than that in the 
$p_{\rm x}$-wave symmetry $\phi_{1{\rm x}}(\k)$. 
 This lifting of degeneracy is very small and these two states are 
nearly degenerate. 
 Therefore, it is expected that the orbital effect plays 
a more important role for the lifting of degeneracy. 
 As will be discussed in \S4.1, the degeneracy between 
$\hat{d}=\alpha p_{\rm x}\hat{x}+p_{\rm y}\hat{y}$ and 
$\hat{d}=\alpha p_{\rm y}\hat{x}-p_{\rm x}\hat{y}$ is lifted by 
the the orbital effect, and the former is stabilized furthermore. 
 According to the analysis on the Ginzburg-Landau 
free energy,~\cite{rf:sigrist,rf:agterberg,rf:yanasepart3} 
the orbital effect on the lifting of this degeneracy is in the second 
order of order parameter, while the paramagnetic effect is beyond the 
forth order.

 The B-phase is stabilized at high magnetic fields. 
 The reflection symmetry with respect to the {\it xy}-plane 
is violated in the B-phase. 
 In comparison to the A-phase, the Zeeman coupling energy is gained 
in the B-phase because the {\it x}-component of $d$-vector is reduced, 
although the spin-orbit coupling term disfavors the B-phase. 
 The A-B phase transition is described by the rotation of $d$-vector 
having the $p_{\rm x}$-orbital component, which is shown in Fig.~3 
schematically. 
 It should be stressed that the $\hat{x}$-component of $d$-vector 
does not vanish even in the B phase owing to the $g_3$-term in 
eq.~(\ref{eq:single-orbital-model}). 
 Because the amplitude of $p_{\rm x}$-wave order parameter increases 
as increasing the magnetic field across the A-B phase transition, 
the superconducting gap becomes more isotropic in the B-phase. 
 In comparison to the C-phase, the B-phase is stabilized at low 
temperatures since the superconducting gap is more isotropic in the 
spin space rather than the non-unitary C-phase. 
 At high fields, the B'-phase where 
$\hat{d}=\alpha p_{\rm y}\hat{x}-p_{\rm x}\hat{y} 
\pm \beta p_{\rm y}\hat{z}$ is nearly degenerate with the B-phase and 
can be stabilized. 
 However, we have ignored this possibility since the B'-phase is 
disfavored by the orbital effect which is discussed in \S4.1.

\begin{figure}[ht]
\begin{center}
\includegraphics[width=7cm]{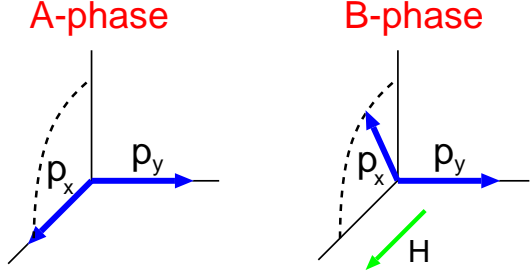}
\caption{
 The schematic figure for the A-B transition. 
 The $d$-vector with $p_{\rm x}$-wave symmetry rotates towards 
{\it z}-axis. 
} 
\end{center}
\end{figure}

 The C-phase is a non-unitary state, {\it i.e.} 
$\hat{d} \times \hat{d^{*}} \ne 0$ where the time-reversal symmetry is 
broken. 
 The C-phase can be stabilized just below $T=T_{\rm c}(H)$ because 
the $d$-vector is described by a linear combination of two-fold degenerate 
state at $T=T_{\rm c}(H)$ as 
$\hat{d}=(\alpha p_{\rm x}\hat{x}+p_{\rm y}\hat{y}) 
\pm {\rm i} (\alpha p_{\rm y}\hat{x}-p_{\rm x}\hat{y})$. 
 In general, a linear combination of the degenerate states is  
stabilized just below $T_{\rm c}(H)$ so as to make the superconducting 
gap most isotropic. 
 The superconducting gap is more isotropic in the spin space 
in the unitary A-phase, while  that is more isotropic in the momentum 
space in the C-phase owing to the factor $p_{\rm x} \pm {\rm i} p_{\rm y}$. 
 As increasing the magnetic field and therefore decreasing $\alpha$, 
the superconducting gap in the A-phase becomes anisotropic while 
that in the C-phase becomes isotropic. 
 Therefore, the C-phase is stabilized at high fields. 
 As decreasing the temperature, the B-phase is stable in comparison to 
the C-phase owing to the non-unitarity in the C-phase. 
 Thus, the C-phase is stabilized at high magnetic fields and 
around $T=T_{\rm c}(H)$.

 The D-phase is the chiral superconducting state where the 
time-reversal-symmetry is broken. 
 It has been expected that this phase is stabilized 
in Sr$_2$RuO$_4$ at $H=0$.~\cite{rf:maeno} 
 However, the D-phase is not stabilized in \Co unless we assume a 
very small spin-orbit coupling as $\lambda=0.05$ in Fig.~2(a).

\begin{figure}[ht]
\begin{center}
\includegraphics[width=7cm]{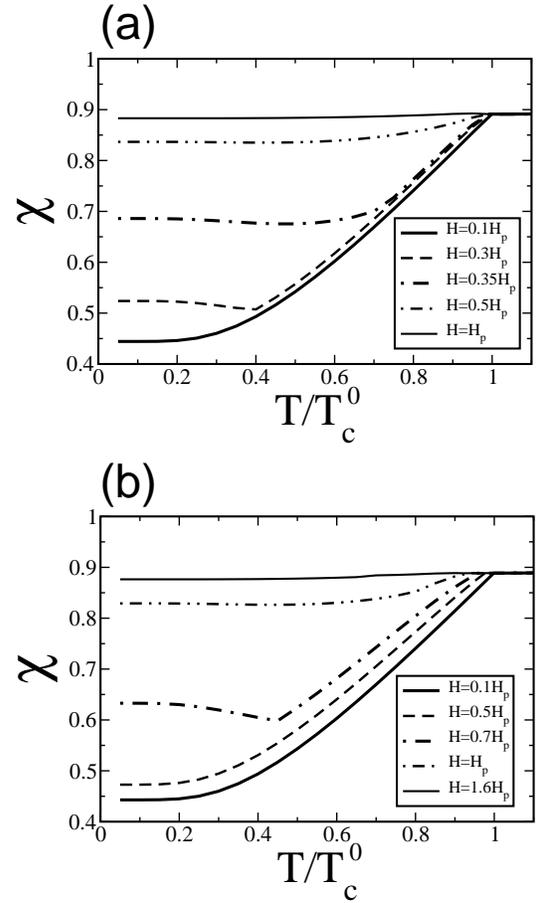}
\caption{The temperature dependence of magnetic susceptibility, 
$\chi=M/H$. 
(a) $\lambda=0.05$ and $H=0.1 H_{\rm p}$, $0.3 H_{\rm p}$, 
$0.35 H_{\rm p}$, $0.5 H_{\rm p}$ and $H_{\rm p}$.
(b) $\lambda=0.17$ and $H=0.1 H_{\rm p}$, $0.5 H_{\rm p}$, 
$0.7 H_{\rm p}$, $H_{\rm p}$ and $1.6 H_{\rm p}$.
} 
\end{center}
\end{figure}

 In the following part, we clarify the physical properties of 
each phase in order to compare with experimental results and 
to suggest future experiments.  
 We first show the temperature dependence of magnetic 
susceptibility $\chi=M/H$ in Fig.~4. 
 It is clearly shown that the magnetic susceptibility is smaller 
than the normal state value in both A-phase and B-phase owing to 
the {\it x}-component of $d$-vector. 
 The decrease of magnetic susceptibility is reduced by increasing the 
magnetic field since the {\it x}-component of $d$-vector decreases so as to 
gain the Zeeman coupling energy. 
 These behaviors are qualitatively consistent with the recent NMR 
measurements in the O-site.~\cite{rf:ihara-o} 
 The magnetic susceptibility is almost temperature independent in the 
C-phase, but it decreases below \Tc when we consider the orbital effect 
as in \S4.1.

\begin{figure}[ht]
\begin{center}
\includegraphics[width=7cm]{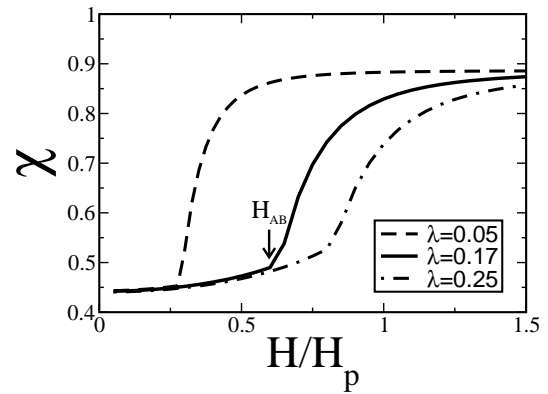}
\caption{
The magnetic field dependence of magnetic susceptibility at 
$T=0.1T_{\rm c}^{0}$. We show the results for 
$\lambda=0.05$, $\lambda=0.17$ and $\lambda=0.25$. 
} 
\end{center}
\end{figure}

 Figure~5 shows the magnetic field dependence of magnetic susceptibility. 
 As shown by the arrow, the magnetic susceptibility rapidly 
increases through the A-B phase transition. 
 This kink will be visible even if we consider the 
vortex state as in \S4.1. 
 Therefore, the detailed measurement on the magnetic field dependence 
of NMR Knight shift will be an interesting experiment. 
  If we assume $\lambda=0.17$ and $T_{\rm c}^{0}=6$K as in Fig.~2(b), 
the A-B transition occurs around $H=0.7 H_{\rm p} \sim 6$T. 
 This magnetic field is experimentally accessible.

\begin{figure}[ht]
\begin{center}
\includegraphics[width=7cm]{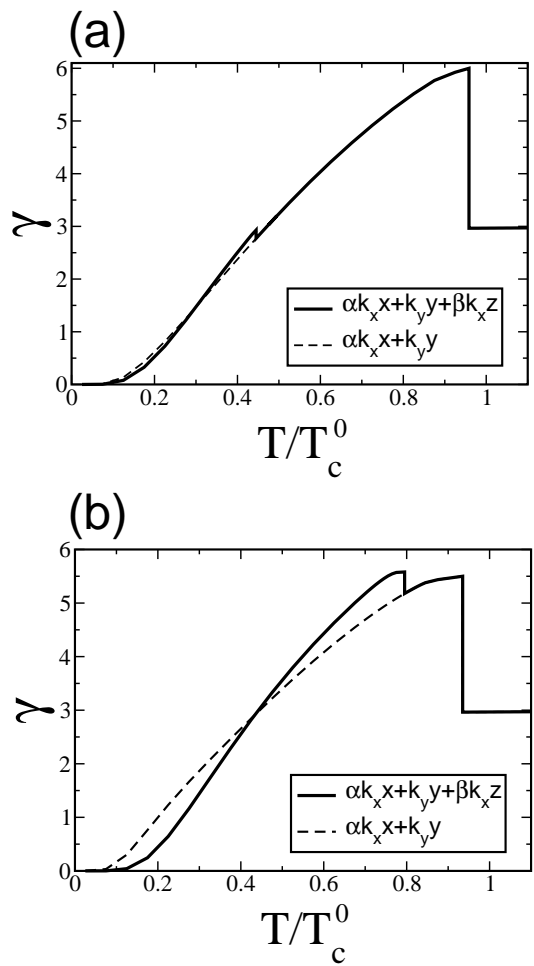}
\caption{
The temperature dependence of specific heat coefficient 
$\gamma=C/T$ (solid line). 
The dashed line shows the results when we assume the A-phase in 
whole temperature range. 
We choose the magnetic field as (a) $H=0.7H_{\rm p}$ and 
(b) $H=H_{\rm p}$. 
} 
\end{center}
\end{figure}

 Since the A-B transition is the second order phase transition, the 
specific heat shows a discontinuity. 
 We calculate the specific heat over temperature $\gamma=C/T$. 
 The typical behavior is shown in Fig.~6(a) where the discontinuity 
in $\gamma$ is not so significant. 
 This is because the entropy is not so different between the A-phase and 
B-phase. It should be noticed that the A-B phase transition is described 
by the rotation of $d$-vector as shown in Fig.~3 and the amplitude of 
order parameter changes only slightly. 
 We note that the broadening of the specific heat occurs in the vortex state. 
 Therefore, it may be difficult to observe such a small discontinuity 
in $\gamma$ experimentally. 
 Indeed, no evidence has been obtained for the multiple phase transition 
from the specific heat measurement in  
\Cof.~\cite{rf:hdyang,rf:lorenz,rf:oeschler} 
 However, this is not incompatible with the existence of 
multiple phase transition. 
 It is worth noting that the specific heat measurement has not detected 
the multiple phase transition in UPt$_3$ at high 
fields.~\cite{rf:ramirez} 
 If we choose the magnetic field so that the second transition occurs near 
$T_{\rm c}(H)$ as in Fig.~6(b), the discontinuity in $\gamma$ becomes 
large. It may be possible to observe the A-B transition in the 
specific heat measurement by adjusting the magnetic field.

\begin{figure}[ht]
\begin{center}
\includegraphics[width=7cm]{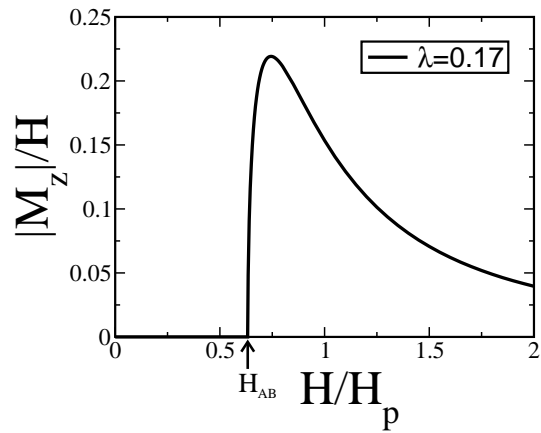}
\caption{
The magnetization along {\it z}-axis over the magnetic field 
along {\it x}-axis. 
The parameters are chosen to be $T=0.1T_{\rm c}^{0}$ and $\lambda=0.17$. 
} 
\end{center}
\end{figure}

 We find another interesting property of the A-B phase transition, 
namely the spontaneous rotation of the principal axis of magnetic 
susceptibility. 
 In the A-phase, the principal axis is the same as the symmetry axis 
of crystal. On the other hand, the principal axis in the B-phase is 
misaligned from the crystal axis because the $d$-vector having 
the $p_{\rm x}$-symmetry lies between the $\hat{x}$ and 
$\hat{z}$ direction. 
 This particular property is illuminated by showing the magnetization 
along the {\it z}-axis, $M_{\rm z}$ under the parallel magnetic field. 
 Although the magnetization $M_{\rm z}$ is zero in the A-phase and in 
the normal state, $M_{\rm z}$ has a finite value in the B-phase 
due to the off-diagonal element of susceptibility tensor. 
 We show the magnetic field dependence of $|M_{\rm z}|/H_{\rm x}$ 
in Fig.~7. 
 The absolute value is shown because the sign of $M_{\rm z}$ 
is different between the $2$-fold degenerate states in B-phase 
(see Table~I). 
 It is shown that the magnetization $|M_{\rm z}|/H_{\rm x}$ shows 
a sharp peak in the B-phase close to the A-B transition. 
 When we increase the magnetic field from the A-B transition line, 
$|M_{\rm z}|/H_{\rm x}$ first increases owing to the increase of 
$\beta$ and then decreases owing to the decrease of $\alpha$. 
 Since the appearance of $|M_{\rm z}|$ is a particular property 
of B-phase, the measurement of this quantity will be an 
interesting experiment. 
 We note that  $|M_{\rm z}|$ is finite also in the C-phase because 
of the non-unitarity. 
 However, the value in the C-phase is in the order of 
$T_{\rm c}^{0}/W$ and therefore much smaller than that in the B-phase.

\begin{figure}[ht]
\begin{center}
\includegraphics[width=6cm]{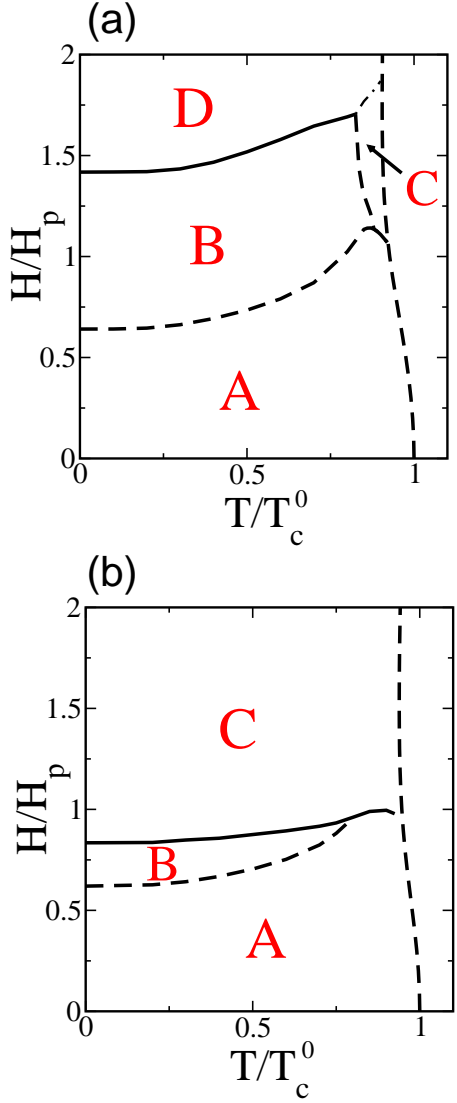}
\caption{
The phase diagram when the value of $g_{1}-g_{5}$ is finite. 
We fix $\lambda=0.17$ and choose $T_{{\rm c}2}=0.825 T_{\rm c}^{0}$ 
($g_{1}-g_{5} < 0$) in (a) 
and $T_{{\rm c}2}=0.905 T_{\rm c}^{0}$ ($g_{1}-g_{5} > 0$) in (b). 
} 
\end{center}
\end{figure}

 Thus far, we have assumed $g_{1}-g_{5}=0$. 
 At the last of this subsection, we discuss the role of the spin-orbit 
coupling through the finite value of $g_{1}-g_{5}$. 
 We fix $\lambda=0.17$ as Fig.~2(b) and choose two values of 
$g_{1}-g_{5}$ so that $T_{{\rm c}2}=0.825 T_{\rm c}^{0}$ and 
$T_{{\rm c}2}=0.905 T_{\rm c}^{0}$. Here, $T_{{\rm c}2}$ is defined as, 
\begin{eqnarray}
\label{eq:mf-tc2} 
1=g_{1} \sum_{\k} \frac{|\phi_{1{\rm x}}(\k)|^{2}}{2 E_{2}(\k)}
\tanh\frac{E_{2}(\k)}{2 T_{{\rm c}2}}. 
\end{eqnarray}
 Since $T_{{\rm c}2}=T_{{\rm c}}^{({\rm z})}=0.865 T_{\rm c}^{0}$ 
for $g_{1}-g_{5}=0$, $g_{1}-g_{5} < 0$ in 
the former case and $g_{1}-g_{5} > 0$ in the latter case. 
 Figure~8 shows the phase diagram in each case.  
 In both cases, the phase diagram in the low field region is similar to 
Fig.~2(b). For instance, the A-B phase transition occurs around 
$H=0.7H_{\rm p}$. 
 On the other hand, the pairing state at high magnetic fields 
is affected by the finite value of $g_{1}-g_{5}$. 
 The D-phase is stabilized when $g_{1}-g_{5} < 0$ while the C-phase is 
stabilized when $g_{1}-g_{5} > 0$. 
 These results are understood by noticing that $g_3$-term 
in eq.~(\ref{eq:single-orbital-model}) is ineffective at high magnetic 
fields since $\alpha \ll 1$. 
 Then, the D-phase is favored when the $g_{5}$-term is large 
while the C-phase is favored when the $g_{1}$-term is large.

 As shown in I,~\cite{rf:yanasepart1} 
the second order term in $\lambda$ stabilizes the 
$P_{\rm xy+}$-state furthermore. 
 This means that $g_{1}-g_{5} > 0$. 
 Since a large value of $g_{1}-g_{5}$ has been assumed in Fig.~8(b), 
the role of $g_{1}-g_{5}$ may be overestimated. 
 We expect that the C-phase is slightly favored by the finite 
value of $g_{1}-g_{5}$ in \Cof.

\begin{figure}[ht]
\begin{center}
\includegraphics[width=7cm]{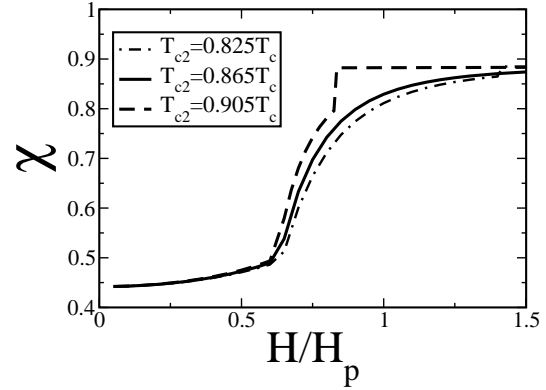}
\caption{
The magnetic field dependence of magnetic susceptibility 
at $T=0.1T_{\rm c}^{0}$ and $\lambda=0.17$. 
We show the results for $g_{1}-g_{5} < 0$, 
$g_{1}-g_{5} = 0$ and $g_{1}-g_{5} > 0$, respectively. 
} 
\end{center}
\end{figure}

 Figure~9 shows the field dependence of magnetic 
susceptibility for $g_{1}-g_{5} \ne 0$. 
 Although the behaviors in the low field region are independent of 
the value of $g_{1}-g_{5}$, the magnetic susceptibility at high fields 
is affected. 
 The discontinuity appears at the B-D and B-C transition which 
are the first order phase transition. 
 The magnetic susceptibility in the C- and D-phases 
does not decrease from the normal state value. 
 We note again that the magnetic susceptibility decreases in the C-phase 
when we take into account the orbital effect as in \S4.1.

\subsection{$F$-wave state}

 The phase diagram in the $f$-wave superconducting state is much 
simpler than the $p$-wave superconducting state. 
 The multiple phase transition is not expected under the magnetic 
field along the plane even if the $F_{\rm xy}$-state is stabilized 
at $H=0$. 
 Among the $2$-fold degeneracy in $F_{\rm xy}$-state, 
$\hat{d}=\alpha f_{2} \hat{x} + f_{1} \hat{y}$ is favored by 
the magnetic field along {\it x}-axis, as shown in Fig.~10 
schematically. 
 Although the magnetic susceptibility slightly decreases owing 
to the $\hat{x}$-component of $d$-vector, the decrease is much less than 
$1\%$ in our estimation 
because $\alpha \ll 1$.~\cite{rf:yanasepart1} 
 Then, the level crossing between the $F_{\rm xy}$- and 
$F_{\rm z}$-state due to the Zeeman coupling energy occurs 
at the extraordinary high magnetic field 
which will be higher than $H_{{\rm c}2}$.

\begin{figure}[ht]
\begin{center}
\includegraphics[width=7cm]{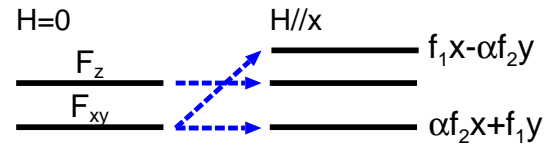}
\caption{
 The schematic figure for the energy level in the $f$-wave 
superconducting state. 
} 
\end{center}
\end{figure}

 If we apply the magnetic field along {\it z}-axis, 
the rotation of $d$-vector occurs when the $F_{\rm z}$-state is 
stabilized at $H=0$. However, it will be difficult to observe 
this phase transition because the magnetic field is very small 
owing to the remarkably small splitting of \Tc between 
$F_{\rm xy}$- and $F_{\rm z}$-states, as shown in I.~\cite{rf:yanasepart1}

 From these discussions, it is expected that the single phase would be 
observed in the $f$-wave superconducting state in both directions of 
magnetic field. 
 We note that almost temperature independent magnetic susceptibility 
expected in the $f$-wave state is incompatible with the results of 
Co-NMR~\cite{rf:kobayashi,rf:yoshimura,rf:ishidaprivate,rf:zhengprivate} 
and O-NMR~\cite{rf:ihara-o}.

\subsection{$P+F$-wave state}

 Finally, we investigate the coexistent $p+f$-wave state. 
 We denote this state as E-phase in Table~I. 
 In the $\Jh$-$n_{\rm e}$ phase diagram in Fig.~1, 
the coexistent state is stabilized in the region denoted as 
$P_{\rm xy+}+F_{\rm z}$. 
 Although such a coexistent state with different orbital symmetry is 
usually realized only accidentally, the $P_{\rm xy+}+F_{\rm z}$-state is 
stabilized in a relatively large parameter region. 
 This is because the $p$-wave and $f$-wave superconductivities 
are nearly degenerate. 
 We have provided a clear explanation on this degeneracy by analyzing 
the orbital character of six hole pockets.~\cite{rf:yanase}

 In order to investigate the $p+f$-wave state in the weak coupling 
approximation, 
we consider the $g_1$-, $g_2$- and $g_3$-terms having 
the $p$-wave symmetry and $g_5$-term having the $f_1$-wave symmetry. 
 Hence, $g_{1}=g_{2}$ due to the symmetry of triangular lattice, 
however $g_{4} \ne g_{5}$ because the $f_1$-wave symmetry is not 
degenerate with the the $f_2$-wave symmetry which is induced the 
$g_4$-term. Hereafter, we ignore the $g_4$-term since the \Tc for 
the $f_2$-wave symmetry is very small.~\cite{rf:yanase}
 For simplicity, we assume the orbital wave functions in the 
$p_{\rm x}$-, $p_{\rm y}$- and $f_1$-symmetry as 
$\phi_{\rm 1x}(\k)=\sqrt{3} 
\sin \frac{\sqrt{3}}{2}k_{\rm x} \cos \frac{1}{2}k_{\rm y}$,
$\phi_{\rm 1y}(\k)=\sin k_{\rm y} + 
\sin \frac{1}{2}k_{\rm y} \cos \frac{\sqrt{3}}{2}k_{\rm x}$ and 
$\phi_{\rm 2y}(\k)=\sin \frac{1}{2}k_{\rm y} 
(\cos \frac{1}{2}k_{\rm y} - \cos \frac{\sqrt{3}}{2}k_{\rm x})$, 
respectively. These wave functions denote the Cooper 
pairing between the nearest neighbor sites. 
 The parameters $g_{1}=g_{2}$, $g_3$ are determined from 
eqs.~(\ref{eq:mf-pxy+}) and (\ref{eq:mf-tc2}) by assuming 
$T_{\rm c2}=0.865 T_{\rm c}^{\rm (xy+)}$ and $T_{\rm c}^{\rm (xy+)}=6$K. 
 We choose the critical temperature in the $f_1$-wave symmetry, 
namely $T_{{\rm c}}^{({\rm z})}$ as a parameter and 
determine $g_5$ by eq.~(\ref{eq:mf-pz}), 
although $T_{{\rm c}}^{({\rm z})}$ can be determined by the microscopic 
parameters, such as $\Jh$ and $n_{\rm e}$.

\begin{figure}[ht]
\begin{center}
\includegraphics[width=6.2cm]{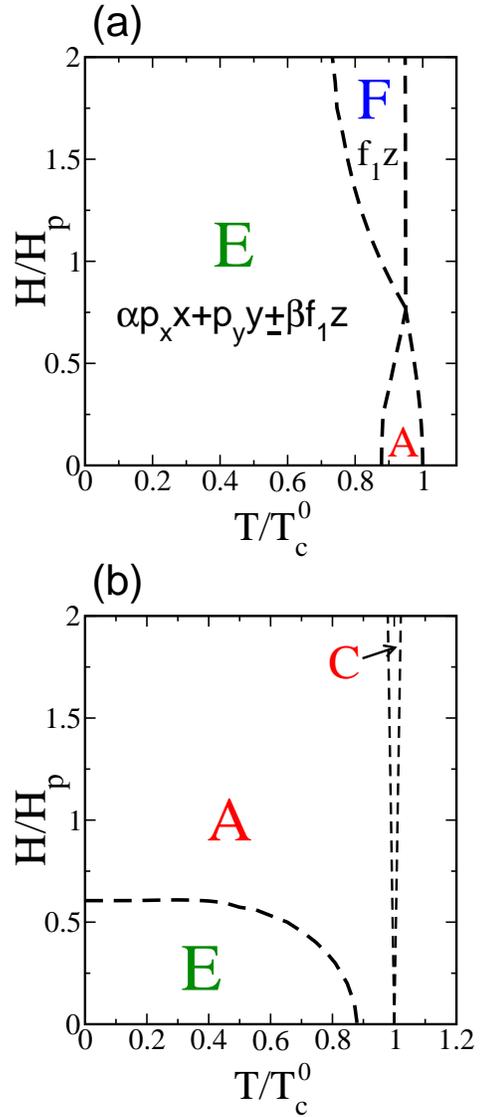}
\caption{
 The $H$-$T$ phase diagram in the $p+f$-wave superconducting state for 
(a) $H \parallel x$ and (b) $H \parallel z$. 
 We choose the parameters, $g_{1}=g_{2}$, $g_3$ and $g_5$, so that 
$T_{{\rm c}2}=0.865 T_{\rm c}^{({\rm xy+})}$ and 
$T_{{\rm c}}^{({\rm z})}=0.95 T_{\rm c}^{({\rm xy+})}$. 
} 
\end{center}
\end{figure}

 We find three kinds of the $H$-$T$ phase diagram 
which are determined by $T_{\rm c}^{({\rm z})}$.  
 These are classified as 
(A) $T_{\rm c}^{({\rm z})} < T_{{\rm c}2} < T_{\rm c}^{({\rm xy+})}$,
(B) $T_{{\rm c}2} < T_{\rm c}^{({\rm z})} < T_{\rm c}^{({\rm xy+})}$ and 
(C) $T_{{\rm c}2} < T_{\rm c}^{({\rm xy+})} < T_{\rm c}^{({\rm z})}$. 
  Since the case (B) is obtained in the largest parameter range 
of two-orbital Hubbard model,~\cite{rf:yanasepart1} 
we assume the case (B). 
 The phase diagram in this case is shown in Fig.~11.

 If we apply the magnetic field along {\it x}-axis, 
there appears a tetracritical point as in Fig.~11(a). 
 The A-phase is stabilized around $T=T_{\rm c}$ at low magnetic fields
since $T_{\rm c}^{({\rm z})} < T_{\rm c}^{({\rm xy+})}$. 
 At high magnetic fields, the F-phase ($F_{\rm z}$-state) is stabilized 
round $T=T_{\rm c}(H)$ since $T_{{\rm c}2} < T_{\rm c}^{({\rm z})}$. 
 In the low temperature region, the E-phase is stabilized in comparison 
to the A-phase and F-phase since the superconducting gap is isotropic. 
 The coexistent E-phase appears below the second transition temperature 
which is lower than $T_{\rm c}^{({\rm z})}$.

 We note that Fig.~11(a) is obtained by neglecting 
the possibility of $F_{\rm xy}$-state. 
 The $F_{\rm xy}$-state can be stabilized  instead of the F-phase 
when the $F_{\rm xy}$-state has higher transition temperature than 
the $F_{\rm z}$-state. 
 In this case, it is sufficient to substitute the $F_{\rm xy}$-state 
for the $F_{\rm z}$-state in the following discussions. 
 Note that the B-phase and D-phase are not stabilized in the case (B) 
since $T_{{\rm c}2} < T_{\rm c}^{({\rm z})}$.

 Fig.~11(b) shows the phase diagram under the magnetic field along 
{\it z}-axis. 
 In this case, we obtain $\alpha=1$ in Table~I 
since the symmetry of point group is not reduced. 
 It is shown that the E-phase disappears at high magnetic fields 
because the magnetic field disfavors the $d$-vector along {\it z}-axis.  
 The C-phase is stabilized around $T=T_{\rm c}(H)$. 
 The degeneracy in the C-phase is lifted because the magnetization 
arising from the non-unitarity couples to the magnetic field. 
 In the present case, 
$\hat{d}=(p_{\rm x} + {\rm i}p_{\rm y}) (\hat{x}-{\rm i}\hat{y})$ 
is favored, where only the down spins are superconducting.  
 This is because the DOS at Fermi level is larger for down spins 
owing to the van-Hove singularity at the Brillouin zone boundary.

\begin{figure}[ht]
\begin{center}
\includegraphics[width=7cm]{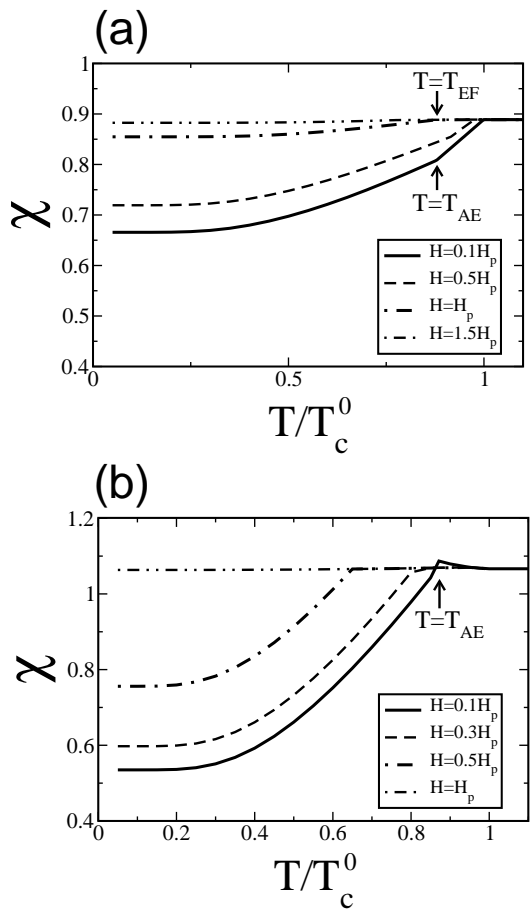}
\caption{
The temperature dependence of magnetic susceptibility 
(a) along {\it x}-axis and (b) along {\it z}-axis. 
We choose the same parameters as in Fig.~11 and the values of 
magnetic field are shown in the legend. 
The arrows show the second phase transition. 
$T_{\rm AE}$ is the transition temperature from A- to E-phase 
at $H=0.1H_{\rm p}$ and $T_{\rm EF}$ is that from E- to F-phase at 
$H=H_{\rm p}$. 
} 
\end{center}
\end{figure}

 One of the characteristic properties in E-phase is the 
decrease of magnetic susceptibility {\it in all directions} 
of magnetic field. 
 This is because the $d$-vector has all components, 
$\hat{x}$, $\hat{y}$ and $\hat{z}$. 
 We show the temperature dependence of magnetic susceptibility 
along {\it x}-axis in Fig.~12(a) and that along {\it z}-axis 
in Fig.~12(b). 
 This feature of E-phase is interesting because 
the decrease of NMR Knight shift along {\it z}-axis has been reported 
by Kobayashi {\it et al.}~\cite{rf:kobayashi} 
 This consistency should be contrasted to the $p$-wave state (\S3.1) and 
$f$-wave state (\S3.2) where the magnetic susceptibility along 
{\it z}-axis does not decrease below \Tcf. 
 We note that the magnetic susceptibility does not decrease in 
the A- and C-phases in Fig.~11(b).

\begin{figure}[ht]
\begin{center}
\includegraphics[width=7cm]{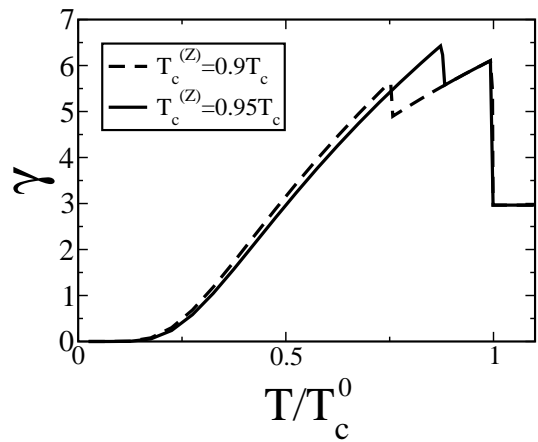}
\caption{
The temperature dependence of the specific heat coefficient 
in the $p+f$-wave state for two values of $T_{{\rm c}}^{({\rm z})}$. 
The solid and dashed lines show the results for 
$T_{{\rm c}}^{({\rm z})}=0.95 T_{\rm c}^{({\rm xy+})}$ and 
$T_{{\rm c}}^{({\rm z})}=0.9 T_{\rm c}^{({\rm xy+})}$, respectively.  
We fix $T_{{\rm c}2}=0.865 T_{\rm c}^{({\rm xy+})}$ and $H=0$. 
} 
\end{center}
\end{figure}

 Another important feature of the $p+f$ coexistent state is the 
existence of double transition at zero magnetic field. 
 This is in sharp contrast to the phase diagram in the $p$-wave state 
(Fig.~2) which shows no double transition at $H=0$. 
 The second transition from A to E-phase induces a clear discontinuity 
of specific heat coefficient $\gamma$ as shown in Fig.~13. 
 This is because the superconducting gap is much more isotropic 
in the E-phase than in the A-phase. 
 The $p$-wave order parameter is remarkably anisotropic 
on the \egf since both $p_{\rm x}$- and $p_{\rm y}$-wave order 
parameters vanish around the K-point owing to the periodicity of 
Burillouin zone. 
 In contrast to that, the $f_1$-wave order parameter 
is almost isotropic on the \egff. 
 Therefore, the entropy is significantly reduced by the appearance of 
$f$-wave order parameter. 
 Although the specific heat has been measured 
at zero magnetic field,~\cite{rf:hdyang,rf:lorenz,rf:oeschler} 
no sign of the second phase transition has been reported up to now. 
 We suggest that an experimental search for this second phase transition 
will be an interesting future issue. 
 
 We have calculated the NMR $1/T_{1}T$ in the $p+f$-wave state, but 
we can see a very weak anomaly at the second phase 
transition.~\cite{rf:yanaseNMR} 
 Therefore, it will be difficult to observe this second phase transition 
by the measurement of NMR $1/T_{1}T$.

\section{Roles of Vortex State and $A_{\rm 1g}$-Orbital}

 In \S3, we have ignored two effects which may 
affect the $H$-$T$ phase diagram. One is the orbital effect which 
induces the vortex state. 
 The other is the role of $a_{\rm 1g}$-orbital. 
 In this section, we investigate these effects within the qualitative 
discussion.

\subsection{Vortex state}

 First, we discuss the vortex state. 
 Most of the unconventional superconductors including \Co are classified 
into type II where the vortex state is stabilized 
at $H_{\rm c1} < H < H_{\rm c2}$. 
 The spatial dependence forming the vortex state is not 
important for the $d$-vector, qualitatively 
when the orbital part of order parameter has one component. 
 The $f$-wave state discussed in \S3.2 is the case. 
 However, the $d$-vector can be affected by the 
orbital effect when the orbital part has multi-component as in the 
$p$-wave and $p+f$-wave states. 
 In these states, the $d$-vector in the vortex state is 
discussed in the following way.

 We discuss the vortex state on the basis of the 
Ginzburg-Landau theory which provides qualitative and clear understandings. 
 In this theory, the free energy is given by the functional of 
order parameters as, 
\begin{eqnarray}
\label{eq:GL}
F=F_{0} + F_{\rm G}, 
\end{eqnarray}
where $F_{0}$ is the uniform term and $F_{\rm G}$ is the gradient term. 
 Here, we define the order parameters $d_{\alpha}^{(\beta)}$ through 
the $d$-vector as 
$\hat{d}=\sum_{\alpha \beta} d_{\alpha}^{(\beta)} \phi_{i \beta}(\k) 
\hat{\alpha}$ where $i=1$ for $\alpha={\rm x,y}$ and 
$i=2$ for $\alpha={\rm z}$. 
 In this subsection, we analyze the GL functional within the quadratic 
order and determine the pairing state at $T=T_{\rm c}(H)$ and 
discuss the $d$-vector below $T_{\rm c}(H)$.

 We first consider the $p$-wave superconductivity.  
 Within the quadratic order of uniform term $F_0$, 
there is a 2-fold degeneracy between 
$(d_{\rm x}^{(y)},d_{\rm y}^{(x)})= 
\Delta_{1}(\frac{\alpha}{\sqrt{1+\alpha^{2}}},-\frac{1}{\sqrt{1+\alpha^{2}}})$
and 
$(d_{\rm x}^{(x)},d_{\rm y}^{(y)})=
\Delta_{2}(\frac{\alpha}{\sqrt{1+\alpha^{2}}},\frac{1}{\sqrt{1+\alpha^{2}}})$.
 These states are obtained from the doubly degenerate 
$P_{\rm xy+}$-state in an adiabatic way and nearly degenerate 
in the A-phase. 
 By taking into account the two component order parameters, namely 
$\Delta_{1}$ and $\Delta_{2}$, the quadratic order term in $F_{0}$ 
is written as 
$F_{0} = a (T-T_{\rm c}^{({\rm xy})}(H)) [|\Delta_{1}|^{2}
+|\Delta_{2}|^{2}]$ where $T_{\rm c}^{({\rm xy})}(H)$ is the 
transition temperature obtained in \S3.1. 
 Note that the A-phase and C-phase are described by a linear 
combination of these two component order parameters. 
 We obtain $(\Delta_{1}, \Delta_{2}) \propto (0,1)$ in the A-phase, 
and $(\Delta_{1}, \Delta_{2}) \propto (\pm {\rm i},1)$ in the C-phase.

 When we consider the magnetic field along {\it x}-axis and 
ignore the spatial dependence in {\it x}-direction, 
the gradient term is given as, 
\begin{eqnarray}
&& \hspace{-8mm}  
F_{\rm G} = K_{1}^{\parallel} \{|D_{\rm y} d_{\rm x}^{(x)}|^{2} 
                         + |D_{\rm y} d_{\rm y}^{(x)}|^{2} \}
           + K_{1}^{\perp}  |D_{\rm y} d_{\rm z}^{(x)}|^{2} 
\nonumber
\\
&& \hspace{-1mm} 
           + K_{2}^{\parallel}|\{|D_{\rm y} d_{\rm x}^{(y)}|^{2} 
                         + |D_{\rm y} d_{\rm y}^{(y)}|^{2} \}
          + K_{2}^{\perp}  |D_{\rm y} d_{\rm z}^{(y)}|^{2} 
\nonumber
\\
&& \hspace{-1mm}       + K_{3}^{\parallel} \{|D_{\rm z} d_{\rm x}^{(x)}|^{2} 
                         + |D_{\rm z} d_{\rm y}^{(x)}|^{2} 
                         + |D_{\rm z} d_{\rm x}^{(y)}|^{2} 
                         + |D_{\rm z} d_{\rm y}^{(y)}|^{2} \}
\nonumber
\\
&& \hspace{-1mm}      + K_{3}^{\perp}  \{|D_{\rm z} d_{\rm z}^{(x)}|^{2} 
                           + |D_{\rm z} d_{\rm z}^{(y)}|^{2} \},  
\label{eq:gradient-term}
\end{eqnarray}
where $D_{j} = \partial_{j} - 2 {\rm i} e A_{j}$ is 
the gauge invariant differential operator.
 The coupling constants are obtained in the clean limit as, 
$K_{1}^{\parallel}$ : $K_{2}^{\parallel}$ : $K_{3}^{\parallel}$ $=$ 
$< |\phi_{1 {\rm x}}(\k)|^{2} v_{\rm y}^{2} >$ : 
$< |\phi_{1 {\rm y}}(\k)|^{2} v_{\rm y}^{2} >$ : 
$< |\phi_{1 {\rm x}}(\k)|^{2} v_{\rm z}^{2} >$ 
and 
$K_{1}^{\perp}$ : $K_{2}^{\perp}$ : $K_{3}^{\perp}$ $=$ 
$< |\phi_{2 {\rm x}}(\k)|^{2} v_{\rm y}^{2} >$ : 
$< |\phi_{2 {\rm y}}(\k)|^{2} v_{\rm y}^{2} >$ : 
$< |\phi_{2 {\rm x}}(\k)|^{2} v_{\rm z}^{2} >$ 
where $v_{\alpha}$ is the Fermi velocity along 
$\alpha$-direction and $< >$ is the average 
on the Fermi surface. 
 We obtain $K_{i}^{\parallel} 
\ne K_{i}^{\perp}$ in general, 
but these are nearly the same values because $\lambda \ll W$. 
 Therefore, we simply assume $K_{i}^{\parallel} = K_{i}^{\perp}=K_{i}$. 
 If the cylindrical Fermi surface and purely $p$-wave order 
parameter are assumed, we obtain $K_{2} = 3 K_{1}$. 
 However, we find that $K_{2} < K_{1}$ 
for the Fermi surface relevant for \Cof.

 It is simply understood that the 2-fold degeneracy between 
$\Delta_{1}$ and $\Delta_{2}$ is lifted by 
taking the gradient term. 
 By taking into account the spatial dependence of order parameters as 
$\Delta_{1}=\Delta_{1}(r)$ and $\Delta_{2}=\Delta_{2}(r)$ and 
solving the linearized GL equation, we obtain two different \Tc as,  
\begin{eqnarray}
\label{eq:tc-split-xy}
&& \hspace{-12mm}
T_{\rm c}^{({\rm xy}1)}(H) = T_{\rm c}^{({\rm xy})}(H) 
- \frac{2e}{a} \sqrt{\frac{\alpha^{2} K_{2}+K_{1}}{1+\alpha^{2}} K_{3}} H, 
\\
&& \hspace{-12mm}
T_{\rm c}^{({\rm xy}2)}(H) = T_{\rm c}^{({\rm xy})}(H) 
- \frac{2e}{a} \sqrt{\frac{\alpha^{2} K_{1}+K_{2}}{1+\alpha^{2}} K_{3}} H. 
\end{eqnarray}
 The former is the critical temperature for the order parameter 
$\Delta_{1}$, while the latter is that for $\Delta_{2}$. 
 Since $K_{2} < K_{1}$ and $\alpha < 1$, we obtain 
\begin{eqnarray}
T_{\rm c}^{({\rm xy}1)}(H) < T_{\rm c}^{({\rm xy}2)}(H). 
\end{eqnarray}
 Thus, the degeneracy between 
$\hat{d}=\alpha p_{\rm x}\hat{x}+p_{\rm y}\hat{y}$ and 
$\hat{d}=\alpha p_{\rm y}\hat{x}-p_{\rm x}\hat{y}$ at $T=T_{\rm c}(H)$ 
is lifted by the orbital effect and the former is stabilized. 
 Therefore, the A-phase is always favored rather than the C-phase at 
$T=T_{\rm c}(H)$ in contrast to Fig.~2(b). 
 The C-phase can be stabilized below $T_{\rm c}(H)$ through the second
order phase transition. 
 This is the most significant role of the orbital effect for the $d$-vector.

 As for the $d$-vector along {\it z}-axis, 
the degeneracy between $\hat{d}=p_{{\rm x}} \hat{z}$ 
and $\hat{d}=p_{{\rm y}} \hat{z}$ 
is lifted by the gradient term in the same way.  
 We obtain the \Tc as, 
\begin{eqnarray}
&& \hspace{-7mm}
T_{\rm c}^{({\rm z}1)}(H) = T_{\rm c}^{({\rm z})}(H) 
-  \frac{2e}{a} \sqrt{K_{1} K_{3}} H, 
\\
&& \hspace{-7mm}
T_{\rm c}^{({\rm z}2)}(H) = T_{\rm c}^{({\rm z})}(H) 
-  \frac{2e}{a} \sqrt{K_{2} K_{3}} H, 
 \end{eqnarray}
where $T_{\rm c}^{({\rm z})}(H)$ is the transition temperature of 
D-phase without including the orbital effect. 
 We obtain 
\begin{eqnarray}
T_{\rm c}^{({\rm z}1)}(H) < T_{\rm c}^{({\rm z}2)}(H), 
\end{eqnarray}
and therefore $\hat{d}=p_{{\rm y}} \hat{z}$ is more favorable than 
$\hat{d}=p_{{\rm x}} \hat{z}$. 
 This splitting of transition temperature is due to the difference of 
coherence length along {\it y}-axis between the $p_{\rm x}$- and 
$p_{\rm y}$-states. 
 The same mechanism for the splitting of \Tc has been discussed 
for Sr$_2$RuO$_4$.~\cite{rf:agterberg2} 
 Then, Agterberg has predicted the double transition at $H > H_{\rm c1}$. 
 However, the experimental observation in Sr$_2$RuO$_4$ 
is incompatible with the theoretical prediction.~\cite{rf:maeno} 
 Recently, Udagawa has proposed another idea 
which resolves this inconsistency.~\cite{rf:udagawa}

 Whether $T_{\rm c}^{({\rm z}2)}(H) < T_{\rm c}^{({\rm xy}2)}(H)$ 
or not depends on the parameters such as $K_{i}$ and 
$T_{\rm c}^{({\rm xy})}(H)-T_{\rm c}^{({\rm z})}(H)$. 
 When we assume $g_{1}-g_{5}=0$ as in Fig.~2, we always obtain 
$T_{\rm c}^{({\rm z}2)}(H) < T_{\rm c}^{({\rm xy}2)}(H)$. 
 Then, the A-phase is always stabilized at $T=T_{\rm c}(H)$.

\begin{figure}[ht]
\begin{center}
\includegraphics[width=7cm]{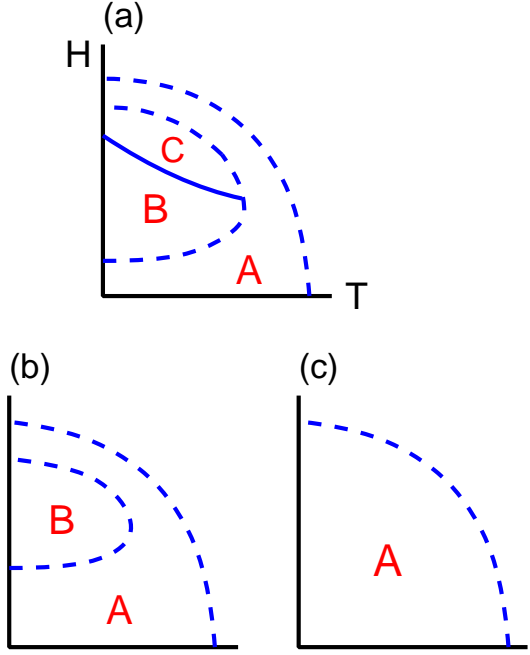}
\caption{
 The schematic figure of $H$-$T$ phase diagram in the $p$-wave 
superconducting state when we take into account the 
vortex state. The magnetic field is applied along {\it x}-axis. 
 The parameter $H_{\rm c2}/H_{\rm p}$ is decreased from (a) to (c). 
} 
\end{center}
\end{figure}

 We show the schematic figure of $H$-$T$ phase diagram in Fig.~14(a-c). 
 The C-phase appears below $T_{\rm c}(H)$ 
when we assume $H_{\rm c2} \gg H_{\rm p}$. 
 Then, the phase diagram is expected as Fig.~14(a). 
 When $H_{\rm c2} \gtrsim H_{\rm p}$, the C-phase may disappear as in 
Fig.~14(b). 
 The B-phase is also destabilized by the gradient term at 
high magnetic fields, however stabilized at low magnetic fields. 
 When $H_{\rm c2}$ is decreased furthermore, the B-phase disappears as 
in Fig.~14(c). 
 In this case, another phase $\hat{d}=p_{y}\hat{z}$ can be stabilized 
at high magnetic fields if we assume $g_{1}-g_{5}<0$. 
 However, this is not expected in our results for the two-orbital Hubbard 
model.~\cite{rf:yanasepart1} 
 Unfortunately, there are two different experiment reports for the
$H_{{\rm c}2}$ in \Cof. 
 The resistivity measurement has reported $H_{{\rm c}2}$ comparable
to $H_{\rm p}$,~\cite{rf:sasaki} while the magnetization measurement
has reported $H_{{\rm c}2}$ much larger than 
$H_{\rm p}$.~\cite{rf:sakurai}

 The orbital effect in the $p+f$-wave state is qualitatively 
different from the $p$-wave state in the sense that 
$K_{i}^{\parallel} \ne K_{i}^{\perp}$. 
 In this case, there are two possibilities of the high field phase 
diagram. 
 When $\frac{\alpha^{2} K_{1}^{\parallel}
+K_{2}^{\parallel}}{1+\alpha^{2}} K_{3}^{\parallel} 
> K_{2}^{\perp} K_{3}^{\perp}$, the phase diagram is qualitatively 
the same as Fig.~11(a). 
 Since the gradient term favors the $f$-wave order parameter, 
the tetracritical point shifts to the lower magnetic field. 
 If we assume $\frac{\alpha^{2} K_{1}^{\parallel}
+K_{2}^{\parallel}}{1+\alpha^{2}} K_{3}^{\parallel} < 
K_{2}^{\perp} K_{3}^{\perp}$, the $f$-wave order parameter is disfavored 
by the gradient term. Then, the E-phase disappears in the high field 
region, and the phase diagram is similar to Fig.~11(b). 
 The microscopic examination of the magnitude relation between 
$\frac{\alpha^{\perp} K_{1}^{\parallel}
+K_{2}^{\parallel}}{1+\alpha^{\perp}} K_{3}^{\parallel}$ 
and $K_{2}^{\perp} K_{3}^{\perp}$ is difficult because 
it is necessary to estimate $K_{3}^{\parallel}$ and $K_{3}^{\perp}$ 
originating from the dispersion relation along {\it z}-axis.

 In this paper, we have focused on the phase transition related
to the $d$-vector and we have not discussed the phase transition 
in the vortex configuration. 
 Because the vortex lattice structure is sensitive to the pairing state, 
the latter phase transition may occur in addition. 
 We have simply ignored this possibility because the magnetic and 
thermodynamic properties are not sensitive to the vortex lattice 
structure. 
 We think that the more sophisticated and quantitative calculation 
on the vortex state is one of the fascinating future issues.

\subsection{Role of $a_{\rm 1g}$-orbital}

 We suggest another possibility of phase transition 
induced by the $a_{\rm 1g}$-orbital. 
 This phase transition originates from the $2$-fold degeneracy in the 
$P_{\rm xy+}$-state which is related to the $p$-wave state (\S3.1) 
and $p+f$-wave state (\S3.3). 
 
 The degeneracy in the $P_{\rm xy+}$-state is due to the $U(1)$ symmetry 
corresponding to the in-plane rotation of spin.~\cite{rf:yanasepart1} 
 This symmetry is broken in the three-orbital Hubbard model including 
the $a_{\rm 1g}$-orbital because the in-plane component of orbital 
moment has the matrix element. 
 Therefore, the degeneracy in the two-orbital Hubbard model will be 
lifted at zero magnetic field. 
 However, it is expected that the lifting of degeneracy is small because 
the $a_{\rm 1g}$-orbital is not so important for the superconducting 
instability.~\cite{rf:mochizuki,rf:yanase} 
 Then, the phase transition can occur in the low magnetic field region.

\begin{figure}[ht]
\begin{center}
\includegraphics[height=4cm]{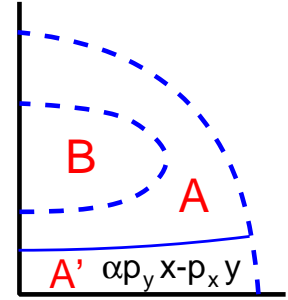}
\caption{
 The schematic figure of $H$-$T$ phase diagram 
when the additional phase transition is induced by 
the $a_{\rm 1g}$-orbital. 
 Although we have assumed the $p$-wave superconducting state, 
qualitatively the same phase transition can occur in the 
$p+f$-wave state. 
} 
\end{center}
\end{figure}

 Let us consider the magnetic field along {\it x}-axis. 
 Then, not only the paramagnetic effect (\S3) 
but also the orbital effect (\S4.1) favor 
$\hat{d}=\alpha p_{\rm x}\hat{x} + p_{\rm y}\hat{y}$, {\it i.e.} A-phase. 
 Therefore, the level crossing occurs under the magnetic field 
when the other $P_{\rm xy+}$-state, {\it i.e.} 
$\hat{d}=p_{\rm y}\hat{x} - p_{\rm x}\hat{y}$, is 
stabilized at $H=0$. 
 We show the schematic phase diagram in Fig.~15 where the additional 
phase transition is assumed.

 The experimental identification for this phase transition 
may be difficult because the magnetic and thermodynamic properties 
are quite similar between these states. 
 However, we note that this phase transition 
is associated with the lattice distortion through the electron-phonon 
coupling. 
 Therefore, this transition may be detected by 
the measurements such as the magnetostriction, thermal expansion and 
ultrasonic attenuation and so on. 
 Actually, the multiple phase transition in UPt$_3$ has been identified 
by these methods.~\cite{rf:muller,rf:dijk}

\section{Summary and Discussion}

 In this paper, we have investigated the multiple superconducting 
phase transition in the two-orbital Hubbard model 
representing the $e'_{\rm g}$-orbitals in \Cof. 
 We have derived the weak coupling model from the linearized 
Dyson-Gorkov equation within the second order perturbation theory 
and solved it within the mean field theory. 

 In order to determine the pairing state under the magnetic field, 
we take into account the atomic spin-orbit coupling term as well as 
the Zeeman coupling term. 
 As a result of these competing effects, the multiple phase transition  
occurs in the superconducting state. 
 We have examined the phase diagram in the $p$-wave 
superconductivity, $f$-wave superconductivity and their coexistent 
state, respectively. 
 The obtained phase diagrams are qualitatively different 
because the role of spin-orbit coupling is quite different between 
the $p$-wave and $f$-wave order parameters. 
 The multiple phase transition occurs under the magnetic field 
along the plane in the $p$-wave superconducting state, 
while that occurs at zero magnetic field in the $p+f$-wave state. 
 However, the multiple phase transition does not occur 
in the $f$-wave state. 
 The obtained results are summarized in Table~II.

\begin{table}[htbp]
 \begin{center}
   \begin{tabular}{|c|ccc|} \hline 
Symmetry & $p$-wave & $f$-wave & $p+f$-wave 
\\\hline 
Double transition & $H \ne 0$ & none & $H=0$ 
\\\hline 
$K(H \perp z)$ & $\bigcirc$ & {\Large $\times$} & $\bigcirc$
\\\hline 
$K(H \parallel z)$ & {\Large $\times$} &  {\Large $\times$} & $\bigcirc$ 
\\\hline 
    \end{tabular}
\caption{Summary of the obtained results. 
 The first column shows the pairing symmetry obtained in the 
multi-orbital Hubbard model. 
 The second column shows the possibility of multiple phase transition. 
 The third and forth columns show the Knight shift 
along the plane and along the $z$-axis, respectively. 
 The symbol $\bigcirc$ ({\Large $\times$}) means that 
the Knight shift decreases (does not decrease) below \Tcf. 
}  
  \end{center}
\end{table}

 The characteristics of each pairing state have been illuminated by
showing the magnetic susceptibility. 
 When we apply the magnetic field along {\it x}-axis, 
the magnetic susceptibility decreases in the $p$-wave state and in the 
$p+f$-wave state, while that does not decrease in the $f$-wave state. 
 The decrease of in-plane Knight shift has been reported by 
the Co-NMR~\cite{rf:kobayashi,rf:yoshimura,rf:ishidaprivate,
rf:zhengprivate} and O-NMR~\cite{rf:ihara-o} which is consistent with 
the $p$-wave state and $p+f$-wave state. 
 When we apply the magnetic field along {\it z}-axis, the 
magnetic susceptibility decreases in the $p+f$ coexistent state, 
while that does not decrease in the $p$-wave and $f$-wave states. 
 A report of {\it c}-axis Knight shift in the 
Co-NMR~\cite{rf:kobayashi} is consistent with the former.

 From these observations, the $p+f$-wave state is absolutely 
consistent with the Knight shift measurements. 
 However, it is still difficult to conclude the pairing state 
which is consistent with the experimental results 
in a comprehensive way. 
 A difficulty of the $p+f$-wave state is that 
the second phase transition has not been observed at zero magnetic field. 
 For instance, 
specific heat~\cite{rf:hdyang,rf:lorenz,rf:oeschler} measurements 
have not detected any anomaly. 
 Another difficulty is the power-law behaviors observed in the 
NMR $1/T_{1}T$ and specific heat. Since the superconducting gap is
almost isotropic in the $p+f$-wave state, the exponential behaviors will 
be observed in the low temperature region.

 Contrary to the $p+f$-wave state, the $p$-wave superconducting state is 
consistent with many experimental results except for the 
{\it c}-axis Knight shift. 
 For instance, the nearly power-law behaviors are expected because 
the superconducting gap is very anisotropic. 
 This is partly because both $p_{\rm x}$- and $p_{\rm y}$-wave 
order parameters vanishes around the K-point near the \egff, 
and furthermore because the nodal quasi-particles exist in the \agff.  
 The $p$-wave state is also consistent with the $\mu$SR 
measurement~\cite{rf:higemoto} which does not detect any 
spontaneous time-reversal-symmetry breaking in contrast to 
Sr$_2$RuO$_4$.~\cite{rf:luke}

 It should be noted that there remain some issues to be resolved 
when we assume the spin singlet pairing. 
 As for the $d$-wave pairing, the $\mu$SR measurement is incompatible 
with the $d_{\rm xy}+{\rm i}d_{\rm x^{2}-y^{2}}$-wave state which is 
expected in the triangular 
lattice.~\cite{rf:baskaran,rf:shastry,rf:lee,rf:ogata} 
 As for the $i$-wave pairing,~\cite{rf:kurokii-wave} 
it seems to be difficult to find a microscopic mechanism 
which induces such a high angular momentum pairing with $T_{\rm c} \sim 5$K.

 From these discussions, we think that further developments in 
the theory and experiment are required to identify the pairing symmetry. 
 The experimental search for the multiple phase diagram is particularly 
interesting because the phase transition related to the $d$-vector is 
a characteristic feature of spin triplet superconductivity. 
 If the multiple phase transition is discovered in the 
superconducting state, that will be a 
conclusive evidence for the spin triplet pairing.

 In order to suggest future experiments, 
we have clarified the nature of multiple phase transition. 
 In the $p$-wave superconductivity, the A-B transition is  
promising to be observed. 
 It is shown that the in-plane Knight shift shows a kink 
at the A-B transition. We find that the magnetization along 
the {\it c}-axis appears in the B-phase under the in-plane magnetic 
field.  This spontaneous rotation of the principle axis of 
susceptibility tensor is a characteristic phenomenon in the $p$-wave 
superconducting state. 

 In order to conclude the $p+f$-wave superconductivity, 
the second phase transition at zero magnetic field should be 
observed experimentally. 
 The accurate measurement of specific heat around $T=$\Tc is highly 
expected because the second transition can be masked by the broadening 
effect arising from the disorder. 

 The measurements on the crystal lattice structure are also interesting 
since the multiple phase transition is associated with the distortion 
of lattice through the electron-phonon coupling. 
 For instance, the magnetostriction, thermal expansion and ultrasonic 
attenuation have played an important role in the studies 
on the heavy fermion superconductors.~\cite{rf:muller,rf:dijk}

 In summary, we have examined the possibilities of multiple phase 
diagram under the parallel magnetic field and suggested some future 
experiments. We hope that the pairing state in \Co will be resolved 
by future experiments on the multiple phase transition.

\section*{Acknowledgments}

 The authors are grateful to Y. Ihara, H. Ikeda, K. Ishida, M. Kato, 
Y. Kitaoka, Y. Kobayashi, C. Michioka, K. Miyake, Y. Ono, 
N. E. Phillips, H. Sakurai, M. Sato, M. Udagawa, Y. J. Uemura, 
K. Yamada, K. Yada and G-q. Zheng for fruitful discussions. 
 Numerical computation in this work was partly carried out 
at the Yukawa Institute Computer Facility. 
 The present work was partly supported by a Grant-In-Aid for Scientific 
Research from the Ministry of Education, Science, Sports and Culture, Japan.

\end{document}